\documentstyle[preprint,aps,eqsecnum]{revtex}
\tighten
\def\a{\alpha}
\def\b{\beta}
\def\g{\gamma}

\def\l{\lambda}
\def\s{\sigma}

\def\Th{\Theta}
\def\o{\over}
\def\rs{r_\sigma}
\def\rt{r_\tau}
\def\rss{r_{\sigma\sigma}}
\def\rtt{r_{\tau\tau}}
\def\ts{t_\sigma}
\def\ta{t_\tau}
\def\tss{t_{\sigma\sigma}}
\def\taa{t_{\tau\tau}}
\def\ps{\phi_\sigma}
\def\pa{\phi_\tau}
\def\pss{\phi_{\sigma\sigma}}
\def\paa{\phi_{\tau\tau}}
\def\tes{\theta_\sigma}
\def\tea{\theta_\tau}
\def\tess{\theta_{\sigma\sigma}}
\def\teaa{\theta_{\tau\tau}}
\newcommand{\tr}{\mbox{tr}}
\begin{document}
\title{\Large{\bf  LECTURES ON STRING  THEORY IN \\ CURVED SPACETIMES}}
\author{{\bf H.J. de Vega$^{(a)}$
 and N. S\'anchez$^{(b)}$}}

\bigskip

\address
{  (a)  Laboratoire de Physique Th\'eorique et Hautes Energies,
Universit\'e Pierre et Marie Curie (Paris VI) et Universit\'e Denis
Diderot (Paris VII),
Tour 16, 1er. \'etage, 4, Place Jussieu
75252 Paris, cedex 05, France.  Laboratoire Associ\'{e} au CNRS
URA280.\\
 (b) Observatoire de Paris, DEMIRM, 61, Avenue de l'Observatoire,
75014 Paris, France.  Laboratoire Associ\'{e} au CNRS URA336,
Observatoire de Paris et \'{E}cole Normale Sup\'{e}rieure.}

\maketitle

\vskip 80pt

\begin{center}

Based on lectures delivered at the THIRD PARIS  COSMOLOGY COLLOQUIUM,

OBSERVATOIRE DE PARIS ~  7 - 9 June  1995,

to appear in the Proceedings edited by  H.J. de Vega and N. S\'anchez,

World Scientific Publ. Co.

\bigskip

and

\bigskip

at the ERICE CHALONGE School, `STRING GRAVITY AND

PHYSICS AT THE PLANCK  SCALE', 8-19 September 1995,

to appear in the Proceedings edited by  N. S\'anchez, Kluwer Publ. Co.

\end{center}
\newpage

\begin{abstract}
Recent progress on string theory in curved spacetimes is
reviewed. The string dynamics in cosmological and black hole spacetimes
is investigated. The different methods available to solve the string equations
of motion and constraints in curved spacetimes are described. That is,
the string perturbation approach,  the null string approach,
the $\tau$-expansion, and the construction of  global solutions
(for instance by inverse scattering methods).

The classical behaviour of strings in FRW and inflationary
spacetimes is now understood in a large extent from
 the various types of explicit  string solutions.
Three different types of behaviour appear:
{\bf unstable, dual} to unstable and {\bf stable}.
For the unstable strings, the energy and size grow  for large    scale
factors $R \to \infty$,  proportional to $R$. For the dual to unstable
strings, the energy and size blow up for $R\to 0$ as $1/R$. For stable
strings,  the energy and proper size are bounded. (In Minkowski
spacetime, all string solutions are of the stable type).

Recent progress on self-consistent solutions to the Einstein equations
for string dominated universes is reviewed. The energy-momentum
tensor for a gas of strings is then considered  as source of the
spacetime geometry and from the above string behaviours  the string
equation of state is {\bf derived}. The  self-consistent
string  solution  exhibits the realistic matter dominated behaviour
$ R \simeq T^{2/3}\; $ for large times and the radiation dominated
 behaviour $ R \simeq T^{1/2}\; $ for early times ($T$ being the cosmic time).

We  report on the {\bf exact integrability} of the string equations
plus the constraints in de
Sitter spacetime that allows to systematically find {\bf exact} string
solutions by soliton methods and the multistring solutions. {\bf Multistring
solutions} are a new feature in curved spacetimes. That is, a single
world-sheet simultaneously describes many different and independent strings.
This phenomenon has no analogue in flat spacetime and follows to the coupling
of the strings with the geometry.

Finally, the string dynamics next and inside a Schwarzschild black hole
is analyzed and their physical properties discussed.
\end{abstract}

\newpage
 \begin{center}
 {\bf CONTENTS}
\end{center}
\begin{itemize}

\item  {\bf I.   Introduction}

\item  {\bf II.  Strings in Curved and Minkowski Spacetimes}.
\begin{description}

\item[A] A brief review on strings in Minkowski spacetime.

\item[B] The string energy-momentum tensor and the string invariant size.

\item[C] Simple String Solutions in Minkowski Spacetime

\end{description}

\item {\bf III.  How to solve the string equations of motion in curved
spacetimes?}
\begin{description}
\item[A] The $\tau$-expansion.

\item[B] Global Solutions.

\end{description}

\item  {\bf IV. String propagation in cosmological spacetimes.}

\begin{description}

\item[A] Strings in cosmological universes: the $\tau$-expansion at
work.

\item[B] The perfect gas of strings.

\end{description}

\item  {\bf V.  Self-consistent string cosmology.}

\begin{description}

\item[A] String Dominated Universes in General Relativity
(no dilaton field).

\item[B]  Thermodynamics of strings in cosmological spacetimes.

\end{description}

\item  {\bf VI.  Effective String  Equations with the String Sources Included.}

\begin{description}

\item[A] Effective String Equations in Cosmological Universes

\item[B]   String driven inflation?

\end{description}

\item  {\bf VII. Multi-Strings and Soliton Methods in de Sitter Universe.}

\item  {\bf VIII.  Strings next to and inside  black holes.}

\begin{description}

\item[A] String Equations of motion in a Schwarzschild Black Hole.

\item[B]   Strings Near the Singularity $r=0$

\item[C] String energy-momentum  and  invariant size near the
singularity.

\item[D] Axisymmetric ring solutions.

\end{description}

\end{itemize}

\section{\bf Introduction}

 The construction of a sensible quantum theory of gravitation is
 probably the greatest challenge in theoretical physics for the end
of this century and most probably for the next century too.

         Another problem (the most often discussed in  connection with gravity
quantization)
  is the one of the renormalizability of the Einstein theory
 (or its various generalizations) when quantized as a local quantum field
 theory. Actually, even deeper conceptual problems arise when one tries to
 combine quantum concepts with General Relativity. For example, statistical
phenomena like Hawking's radiation arise
 when free fields are quantized in black-hole backgrounds. This
points out  a lack of quantum
coherence  even keeping the gravitational field classical.

It may be very
 well that a quantum theory of gravitation needs new concepts and ideas.
 Of course, this future theory must have the today's General Relativity and
 Quantum Mechanics (and QFT) as limiting cases. In some sense, what everybody
  is doing in this domain (including string theories approach)
may be something analogous to the developpment of the old quantum theory
in the 10's of this century. Namely, people at that time imposed quantization
conditions (the Bohr-Sommerfeld conditions) to hamiltonian mechanics
but keeping the  concepts of  classical mechanics.

         The main drawback to develop a quantum theory of gravitation is
 clearly the {\bf total lack of experimental guides} for the theoretical
 developpment. Just from  dimensional reasons, physical effects combining
 gravitation and quantum mechanics are relevant only at energies of the
 order of  $M_{Planck}  =  \sqrt{ \hbar c / G } =  1.22 \,  10^{16} $Tev.
 Such energies were available in the Universe at times $ t < t_{Planck}
 =  5.4 \, 10^{-44} $sec. Anyway, as a question of principle,
 the construction of a quantum theory of gravitation
 is a problem of fundamental
  relevance for theoretical physics . In addition, one cannot rule
 out completely the possibility of some ``low energy'' ($E \ll  M_{Planck}$)
 physical effect that could be experimentally tested. One may speculate
about effects analogous to  the presence of magnetic monopoles in some grand
unified theories. [Monopoles can be detected by low energy experiments
in spite of their large mass].

 Let us now see what are the consequences of Heisenberg's principle
 in quantum mechanics combined with the notion of gravitational
 (Schwarzschild) radius in General Relativity.
         Assume we make two measurements at a very small distance $\Delta x$ .
  Then,
 $$
 \Delta p  \sim \Delta E  \sim  1 / \Delta x
 $$
 where we set  $\hbar = c = 1$ . For sufficiently large  $\Delta E$,
 particles with masses $m \sim  1/\Delta x$ will be produced.
 The gravitational radius of such particles are of the order
 $$
                 R_G \sim G m \sim {(l_{Planck})^2 \o {\Delta x}}
 $$
 where  $l_{Planck}  \sim 10^{-33}$ cm. Now, General Relativity allows
 measures at a distance  $\Delta x$ , provided
 $$
  \Delta x  > R_G  \quad \to \quad  \Delta x   >
{(l_{Planck})^2 \o{ \Delta x}}\qquad .
 $$
 That is,
 \begin{equation}\label{colap}
                  \Delta x  > l_{Planck} \quad \mbox{and}
 \quad  m   <   M_{Planck}
\end{equation}
 This means that no measurements can be made at distances smaller than the
 Planck length and that no particle can be heavier than  $M_{Planck}$ .
  This is a simple consequence of relativistic quantum mechanics combined
  with General Relativity. In addition, the notion of locality and hence
 of spacetime becomes meaningless at the Planck scale. Notice that the
 equality in eq.(\ref{colap}) means  that the Compton length equals
 the Schwarzschild radius of a particle.
   Since  $M_{Planck}$ is the heaviest possible particle scale, a theory valid
 there (necessarily involving quantum gravitation) will also be valid at any
lower
 energy scale. One may ignore higher energy phenomena in a low energy theory,
  but not the opposite. In other words, a theory of
quantum gravity  will be a `theory of everything'.
 We think that this is the {\bf key point} on the quantization of gravity.
 A theory that holds till the Planck scale must describe {\bf all} what happens
 at lower energies including all known particle physics as well as what
  we do not know yet (that is, beyond the standard model) \cite{erice}.
 Notice that this conclusion is totally
 independent of the use or not of string models.
  A direct important consequence of this conclusion, is that it may  not
  make physical sense to quantize {\bf pure gravity}. A physically
 sensible quantum theory cannot contain only gravitons. To give an example,
 a theoretical prediction for graviton-graviton scattering at energies of
 the order of $M_{Planck}$   must include all particles produced in a real
 experiment. That is, in practice, {\bf all} existing particles in nature,
since
 gravity couples to all matter.

 In conclusion : a consistent quantum theory of gravitation must be
  a finite theory \cite{erice}
and must include all other interactions. That is, it must be
  a theory of everything (TOE). This is a very ambitious project. In particular
  it needs the understanding of the present desert between 1 and $10^{16}$ TeV.
         There is an additional dimensional argument about the inference
 Quantum Theory of Gravitation $\to$  TOE. There are only three dimensional
 physical magnitudes in nature: length, energy and time and correspondingly
 only three dimensional constants in nature: $c, \hbar$ and $G$. All
other physical constants like  $\a  = 1/137,04... ,
M_{proton}/m_{electron}, \, \theta_{WS}, \ldots $
 etc. are pure numbers and they must be calculable in a TOE.
 This is a formidable, but extremely appealing problem.
         From the theoretical side, the {\bf only serious candidate} for a TOE
  is at present string theory. This is why we think that strings desserve a
 special attention in order to quantize gravity.

String theory is therefore an appropriate arena to work out the quantization
of gravity consistently. It  provides an unified theory of all
interactions overcoming at the same time the nonrenormalizable
character of quantum fields theories of gravity.

         As a first step in the understanding of quantum gravitational
 phenomena in a string framework, we started in 1987 a programme of string
 quantization on curved spacetimes \cite{dvs87,agn}. The investigation of
strings in curved spacetimes is currently the best framework to study the
physics of gravitation in the context of string theory, in spite of
its limitations. First, the use of a continuous Riemanian manifold to
describe the spacetime cannot be valid at scales of the order of $l_{Planck}$.
More important, gravitational backgrounds effectively provide classical
or semiclasical descriptions even if the matter backreaction to the
geometry is included through semiclassical Einstein equations (or
stringy corrected Einstein equations) by inserting the expectation
value of the string energy-momentum tensor in the r.h.s. One would
want a full quantum treatment for matter and geometry. However, to find
a formulation  of string theory going beyond the use of classical backgrounds
 is a very difficult (but  fundamental) problem.  One would like
 to derive the spacetime geometry as a classical and low energy
  ($ \ll M_{Planck}$) limit from the solution of (quantum) string theory.

After a short introduction on strings in Minkowski and curved
spacetimes, we focus on strings in cosmological spacetimes.

Substantial results were achieved in this field since
1992. The classical behaviour of strings in FRW and inflationary
spacetimes is now understood in a large extent\cite{cos}. This understanding
followed the finding of various types of exact, asymptotic  and
numerical string solutions in  FRW and
inflationary spacetimes\cite{dms} -\cite{din}. For inflationary
spacetimes, the exact
integrability of the string propagation equations plus the string
constraints in de Sitter spacetime \cite{prd} is indeed an
important help.
This allowed to  systematically find {\bf exact} string solutions  by soliton
methods using the linear system associated to the problem
(the so-called dressing method in soliton theory)
and the multistring solutions \cite{dms} -\cite{igor}.

  In summary, three different types of
behaviour are exhibited by the string solutions in cosmological
spacetimes: {\bf unstable, dual} to unstable and {\bf stable}.
For the unstable strings, the energy and size grow  for large    scale
factors $R \to \infty$,  proportional to $R$. For the dual to unstable
strings, the energy and size blow up for $R\to 0$ as $1/R$. For stable
strings, the energy and proper size are bounded. (In Minkowski
spacetime, all string solutions are of the stable type).  The equation
of state for these string behaviours take the form
\begin{itemize}
\item (i) {\bf unstable}  for $ R \to \infty
 \;   p_u =  -E_u/(D-1) < 0 $
\item (ii) {\bf dual to unstable}  for $ R \to 0
 , \; p_d = E_d/(D-1) > 0  $ .
\item (iii) {\bf stable} for $ R \to \infty ,
 \;  p_s = 0 $ .
\end{itemize}
Here $E_u$ and  $E_d$ stand for the corresponding string energies and
$D-1$  for the number of spatial dimensions where the string
solutions lives. For example, $d-1 = 1$ for a straight string,
 $d-1 = 2$ for a ring string, etc.

As we see above,
the dual to unstable string behavior leads to the same equation of
state than radiation   (massless particles or hot matter). The stable
string behavior leads to the  equation of
state of massive particles (dust or cold matter). The unstable string behavior
is a purely `stringy' phenomenon. The fact that in entails a negative
pressure is however  physically acceptable.
For a gas of strings, the unstable string behaviour dominates in
inflationary universes when  $ R \to \infty$ and the   dual to
unstable string behavior dominates for  $ R \to 0 $.

The unstable strings
correspond to the critical case of the so called {\it coasting universe}
\cite{ell,tur}. That is, classical strings provide a {\it concrete}
realization of such cosmological models.
The `unstable' behaviour is called  `string stretching'
in the cosmic string literature \cite{twbk,vil}.

It must be stressed that while  time evolves, a {\bf given} string
solution may exhibit two and even three of the above regimes one after
the other (see sec. III).
 Intermediate behaviours are also observed in ring solutions
\cite{dls,din}. That is,
$$
P = (\g - 1)\; E \quad {\rm with~} -{1 \o {D-1}}< \g - 1< +{1 \o {D-1}}\; .
$$

We also report here on the exact
integrability of the string  equations plus the constraints in de
Sitter spacetime  which  allows to  systematically find {\bf
exact} string solutions  by soliton methods  and the multistring solutions.
 {\bf Multistring solutions} are a new  feature  in curved spacetimes.
  That is,  a single world-sheet  simultaneously describes
many   different and independent strings. This phenomenon has no
analogue in flat spacetime and  appears as a
consequence of the coupling of the strings with the spacetime geometry.

The world-sheet time $\tau$ turns out to be an multi-valued
function of the target string time $X^0$ (which can be
the cosmic time $T$, the conformal time $\eta$ or for de Sitter
universes it can be the hyperboloid time $q^0$).
Each branch of $\tau$ as a
function of $X^0$ corresponds to a different string. In flat spacetime,
multiple string solutions are necessarily described by multiple
world-sheets. Here, a single world-sheet describes  one string,
several strings or even an infinite number of
different and independent strings as a consequence of the coupling with the
spacetime geometry. These strings do not interact among themselves; all the
interaction is with the curved spacetime.  One can decide  to study
separately each of them (they are all different)
or consider all the infinite strings together.

Of course, from our multistring solution, one {\it could} just
choose only one interval in $ \tau $ (or a subset of intervals  in $
\tau $) and describe just one string (or several). This will be just a
{\bf truncation} of the solution.

The really remarkably fact
is that all these infinitely many strings come {\bf naturally together}
when solving the string equations in de Sitter spacetime as we did in
refs. \cite{dms} -\cite{dls}.

Here, interaction among the strings (like splitting and merging) is neglected,
the only interaction is with the curved background.

The multistring property appears associated to the presence of a cosmological
constant (whatever be its sign) \cite{bhln}.
Multistring solutions have not been found in
black-hole backgrounds (without cosmological constant).
More recently, new classes of dynamical and stationary
multistring solutions in curved spacetimes have been found and classified
and their physical properties analyzed \cite{bhln}.
Multistrings has been found  for all inflationary spacetimes  \cite{multi}
but not in FRW universes.

The study of string propagation in curved spacetimes provide essential
clues about the physics in this context but is clearly not the end of
the story. The next step  beyond the investigation of {\bf test}
strings, consist in finding {\bf self-consistently} the geometry from
the strings as matter sources for the Einstein equations
or better the string effective equations (beta functions).
This goal is achieved in ref.\cite{cos} for cosmological spacetimes at the
classical level. Namely, we used  the energy-momentum tensor for a gas
of strings as source for the Einstein equations and we solved them
self-consistently.

To write the string equation of state we used the behaviour of the
string solutions in cosmological spacetimes.
Strings continuously evolve from one type of behaviour to another, as
is explicitly shown by our solutions \cite{prd} -\cite{dls}. For
intermediate values of $ R $, the  equation of state for gas of free strings
 is clearly complicated but a formula of the type:
\begin{equation}
\rho = \left( u_R \; R + {{d} \over R} + s \right) {1 \over
{R^{D-1}}} \label{rogenI}
\end{equation}
 where
\begin{eqnarray}
\lim_{R\to\infty} u_R = \cases{ 0 \quad & {\rm FRW } \cr
  u_{\infty} \neq 0 & {\rm Inflationary } \cr}
\end{eqnarray}
 This equation of state is qualitatively
correct for all $ R $ and becomes exact for $ R \to 0 $ and $ R \to
\infty $ . The parameters
$u_R , d$ and $ s $ are positive constants and the $u_R$
varies smoothly with $R$.

The pressure associated to the energy density (\ref{rogenI}) takes then
the form
\begin{equation}
p  = {1 \over {D-1}} \left( {d \over R} -  u_R \; R\right) {1 \over
{R^{D-1}}} \label{pgenI}
\end{equation}

Inserting  this source into the Einstein-Friedman equations leads to a
self-consistent solution  for  string dominated universes (see sec. VI)
\cite{cos}.
This solution exhibits the realistic matter dominated behaviour
$ R \simeq T^{2/(D-1)}\; $ for large times and the radiation dominated
 behaviour $ R \simeq T^{2/D}\; $ for early times.

For the sake of completeness we analyze in sec. IV
 the effective string equations \cite{cos}.
These equations have been extensively treated in the
literature \cite{eqef} and they are not our central aim.

It must be noticed that there is no satisfactory
derivation of inflation in the context of the effective string equations.
 This does not mean that string
theory is not compatible with inflation, but that the effective string
action approach {\it is not enough} to describe inflation. The
effective string equations are a low energy field theory approximation
to string theory containing only the {\it massless} string modes.
The vacuum energy scales to start inflation are typically of the order
of the Planck mass where the effective string action approximation
breaks down. One must also consider the {\it massive} string modes (which
are absent from  the effective string action) in order to properly get
the cosmological condensate yielding inflation.
De Sitter inflation does not emerge as a solution of the
the effective string equations.

In conclusion, the effective
string action (whatever be the dilaton, its potential and the
central charge term) is not the appropriate framework in which to
address the question of string driven inflation.

Early cosmology (at the Planck time) is probably the best place to test
string theory. In one hand the quantum treatment of gravity is
unavoidable at such scales
and in the other hand, observable cosmological consequences are
derivable from the inflationary stage.
The natural gravitational background is  an
inflationary universe as, for instance,
 de Sitter spacetime. Such geometries are not
string vacua. This means that conformal and Weyl symmetries are broken at
the quantum level. In order to quantize  consistently  strings in such
case, one must enlarge the physical phase space including, in
particular, the  factor  $\exp\phi(\s,\tau)$
in the world-sheet metric [see eq.(\ref{liou})]. This is a very
interesting and open problem. Physically, the origin of such
difficulties in quantum string cosmology comes from the fact that one
is not dealing with an {\bf empty} universe since a cosmological spacetime
 necessarily contains matter. In the other hand, conformal
field theory techniques are till now only adapted to backgrounds for which the
beta functions are identically zero, i. e. sourceless geometries.
A (quantum) string theory treatment of early cosmology necessarily
implies {\bf excited} states, not just string vacua. This problem is
completely open today.

The outline of these  lectures is as follows. Section II
presents an introduction to strings in curved spacetimes including
basic notions on classical and quantum strings in Minkowski spacetime
and introducing the main physical string magnitudes: energy-momentum
and invariant string size.

Section III deals with the several methods of resolution of the
string propagation in curved spacetimes.
(In sections III.A and  III.B we treat the perturbative approaches,
the $\tau$-expansion and the global solutions.

Section V deals with strings in cosmological spacetimes, the
 $\tau$-expansion at work there and we present the perfect gas of
strings as a model for string matter.

In section V we treat
self-consistent string cosmology including the string equations
of state. (Section V.A deals with
general relativity, V.B with the  string thermodynamics).

Section VI discuss  the effective (beta functions) string equations
in the cosmological perspective and the search of inflationary solutions.

In sec. VII, we briefly review the systematic construction of
string solutions in de Sitter universe {\it via} soliton methods and
the new feature of multistring solutions.

Section VIII contains the string dynamics next and inside
Schwarzschild black-holes,
the string behaviour near the $r=0$ singularity and their physical properties.

\section{\bf  Strings in Curved and Minkowski Spacetimes.}

 Let us consider bosonic strings (open or closed) propagating in a
curved D-dimensional spacetime defined by a metric $G_{AB}(X),
0 \leq A,B \leq D-1$.
  The action can be written as
 \begin{equation}\label{accion}
    S  = {{1}\o{2 \pi \a'}} \int d\s d\tau \sqrt{g}\,  g_{\a\b}(\s,\tau) \;
 G_{AB}(X) \,
 \partial^{\a}X^A(\s,\tau) \, \partial^{\b}X^B(\s,\tau)
 \end{equation}
 Here  $ g_{\a\b}(\s,\tau)$  ( $0 \leq \a, \b \leq 1$ ) is the metric in the
 worldsheet, $\a'$ stands for the string tension. As in flat spacetime, $\a'
 \simeq (M_{Planck})^{-2} \simeq ( l_{Planck})^2$  fixes the scale in the
theory.
  There are no other free parameters like coupling constants in string theory.

We will start considering given gravitational backgrounds  $G_{AB}(X)$.
That is, we start to investigate {\em test} strings propagating on a
given spacetime. In section IV,
the back reaction problem will be studied. That is, how the strings
may act as source of the geometry.

String propagation in massless backgrounds other
than gravitational (dilaton, antisymmetric tensor) can be investigated
analogously.

 The string action (\ref{accion}) classically enjoys Weyl invariance
 on the world sheet
 \begin{equation}\label{weyl}
  g_{\a\b}(\s,\tau) \to \l(\s,\tau) \,  g_{\a\b}(\s,\tau)
 \end{equation}
 plus the reparametrization invariance
 \begin{equation}\label{reparam}
 \s \to  \s'  =  f(\s,\tau) \qquad , \qquad \tau \to \tau'   =  g(\s,\tau)
 \end{equation}
Here $ \l(\s,\tau),  f(\s,\tau)$ and $  g(\s,\tau)$ are arbitrary functions.

         The dynamical variables being here the string coordinates
  $X_A(\s,\tau)$ , ($0 \leq A \leq D-1$) and the world-sheet metric
 $ g_{\a\b}(\s,\tau)$ .

Extremizing  $S$  with respect to them yields the classical
equations of motion:
 \begin{eqnarray}\label{movi}
 \partial^{\a}[\sqrt g\, G_{AB}(X) \; \partial_{\a}X^{B}(\s,\tau)]  &=&
 \frac{1}{2}\, \sqrt g\; \partial_{A}G_{CD}(X) \,  \partial_{\a}X^{C}(\s,\tau)
  \, \partial^{\a}X^{D}(\s,\tau) \\
   & & 0 \le A \le D-1 \cr \label{vincu} \cr
 T_{\a\b} ~ &\equiv&  ~ G_{AB}(X)[ \, \partial_{\a}X^{A}(\s,\tau) \,
  \partial_{\b}X^{B}(\s,\tau) \cr
   & -&   \frac{1}{2} \, g_{\a\b}(\s,\tau) \,  \partial_{\g}X^{A}(\s,\tau) \,
 \partial^{\g}X^{B}(\s,\tau)\, ]= 0 ~~,
  \quad 0 \le \a , \beta \le 1.
\end{eqnarray}
 Eqs. (\ref{vincu})  contain only first derivatives and are therefore a set
  of constraints. Classically, we can always use the reparametrization
 freedom (\ref{reparam}) to recast the world-sheet metric on diagonal form
 \begin{equation}\label{liou}
  g_{\a\b}(\s,\tau)  =  \exp[ \phi(\s,\tau) ]\, \, {\rm diag}( -1, +1)
\end{equation}
 In this conformal gauge, eqs. (\ref{movi}) -  (\ref{vincu}) take the
simpler form:
 \begin{eqnarray}\label{conouno}
 \partial_{-+}X^{A}(\s,\tau)  +   \Gamma^{A}_{BC}(X)\,  \partial_{+}
  X^{B}(\s,\tau) \, \partial_{-}X^{C}(\s,\tau) =  0~, \quad
   0 \le A \le D-1 ,
\end{eqnarray}
\begin{eqnarray} \label{conodos}
         T_{\pm\pm} \equiv G_{AB}(X) \, \partial_{\pm}X^{A}(\s,\tau) \,
\partial_{\pm}X^{B}(\s,\tau)\equiv 0\, , \quad  T_{+-} \equiv T_{-+} \equiv 0
  \end{eqnarray}
 where we introduce light-cone variables  $x_{\pm} \equiv \s \pm \tau $
   on the world-sheet and where  $\Gamma^{A}_{BC}(X)$  stand for the
 connections (Christoffel symbols) associated to the metric  $ G_{AB}(X)$.

Notice that these equations in the conformal gauge are still invariant
under the conformal reparametrizations:
 \begin{equation}\label{conftr}
 \s +  \tau \to  \s'  +  \tau' =  f(\s+ \tau) \qquad , \qquad
\s - \tau \to \s' - \tau'   =g(\s-\tau)
 \end{equation}
Here $f(x)$ and  $g(x)$ are arbitrary functions.

The string boundary conditions in curved spacetimes are identical to
 those in Minkowski spacetime. That is,
 \begin{eqnarray}\label{condc}
  X^{A}(\s + 2 \pi,\tau) = \,   X^{A}(\s,\tau) \quad & {\rm closed
\,strings} \cr \cr
    \partial_{\s}X^{A}(0,\tau)=\, \partial_{\s} X^{A}(\pi,\tau) =  \,0
 \quad &  {\rm open \,strings. }
 \end{eqnarray}

\subsection{A brief review on strings in Minkowski spacetime}

         In flat spacetime eqs.(\ref{conouno})  become linear
\begin{equation}\label{dalam}
 \partial_{-+}X^{A}(\s,\tau)=0~, \quad 0 \le A \le D-1 ,
\end{equation}
and one can solve them
 explicitly as well as the quadratic constraint (\ref{conodos}) [see
below]:
\begin{equation}\label{vinm}
 \left[ \partial_{\pm}X^{0}(\s,\tau) \right]^2 - \sum_{j=1}^{D-1}
\left[ \partial_{\pm}X^{j}(\s,\tau) \right]^2 = 0
\end{equation}

The solution of eqs.(\ref{dalam})   is usually written for  closed strings as
\begin{equation}  \label{solM}
         X^{A}(\s,\tau)  =  q^A + 2 p^A \a' \tau + i \sqrt{\a'} \sum_{n \neq 0}
   \frac{1}{n}\{ \a^{A}_{n} \exp[in(\s - \tau)]
         + \tilde  \a^{A}_{n} \exp[-in(\s + \tau )] \}
 \label{solm}
\end{equation}
 where  $q^A$  and  $p^A$  stand for the string center of mass position
  and momentum and  $ \a^{A}_{n}$  and $\tilde \a^{A}_{n}$  describe the right
 and left oscillator modes of the string, respectively. Since the
string coordinates are real,
$$
 {\bar \a}^{A}_{n} =  \a^{A}_{-n} \quad, \quad  {\bar  {\tilde
\a}}^{A}_{n} =
{\tilde \a}^{A}_{-n}
$$
This resolution is no more
 possible in general for curved spacetime where the equations of motion
 (\ref{conouno})  are non-linear. In that case, right and left movers
 interact with themselves and with each other.

In Minkowski spacetime we can also write the solution of the string equations
of motion (\ref{dalam}) in the  form
\begin{equation}
         X^{A}(\s,\tau)  = l^A (\s + \tau) + r^A (\s - \tau)
\end{equation}
where $l^A (x)$ and  $r^A (x)$ are arbitrary functions. Now, making an
appropriate  conformal transformation (\ref{conftr}) we can turn any
of the string coordinates $ X^{A}(\s,\tau)$ (but only one of them)
into a constant times $\tau$. The most convenient choice is the
light-cone gauge where
\begin{equation}\label{gauge}
U \equiv X^0 - X^1 =  2 \, p^U \a' \tau .
\end{equation}
That is, there are no string oscillations along the $U$ direction in
the light-cone gauge. We have still to impose the constraints
(\ref{vinm}). In this gauge they take the form
\begin{equation} \label{V}
 \pm 2 \a' p^U\,
\partial_{\pm}V(\s,\tau)= \sum_{j=2}^{D-1}
\left[ \partial_{\pm}X^{j}(\s,\tau) \right]^2
\end{equation}
where $V \equiv X^0 + X^1$. This shows that $V$ is not an independent
dynamical variable since it expresses in terms of the transverse
coordinates $X^2, \ldots , X^{D-1}$. Only $q^V$ is an independent
quantity.

The physical picture of a string propagating in Minkowski spacetime
clearly emerges in the light-cone gauge. The gauge condition
(\ref{gauge}) tells us that the string `time' $\tau$ is just
proportional to the physical null time $U$. Eqs.(\ref{solM}) shows
that the string moves as a whole with constant speed while it
oscillates around its center of mass. The oscillation frequencies are
all integers multiples of the basic one. The string thus possess an
infinite number of normal modes; $   \a^{A}_{n}, \tilde  \a^{A}_{n}$
classically describe their oscillation amplitudes. Only the modes in
the direction of the transverse coordinates  $X^2, \ldots , X^{D-1}$
are physical. This is intuitively right, since a longitudinal or a
temporal oscillation of a string is meaningless. In summary, the
string in Minkowski spacetime behaves as an extended and composite
relativistic object formed by a
$2(D-2)$-infinite set  of harmonic oscillators.

Integrating eq.(\ref{V}) on $\s$ from $0$ to $2\pi$ and inserting
eq. (\ref{solM}) yields the classical string mass formula:
\begin{equation}\label{masac}
m^2 \equiv  p^U p^V - \sum_{j=2}^{D-1} (p^j)^2 = {1 \o {\a'}}
\sum_{j=2}^{D-1}  \sum_{n=1}^{_\infty}
 \left[   \a^{j}_{n} \, \a^{j}_{-n} + {\tilde  \a^{j}_{n}} \,
{\tilde  \a^{j}_{-n}} \right]
\end{equation}
We explicitly see how the  mass of a string depends on its excitation
state. The classical string spectrum is continuous as we read from
eq.(\ref{masac}). It starts at $m^2 = 0$ for an unexcited string:
$\a^{j}_{n} = \tilde  \a^{j}_{n} = 0 $ for all  $ n$ and $j$.

The independent string variables are:

the transverse amplitudes $\{  \a^{j}_{n},  {\tilde  \a^{j}_{n}},
n\varepsilon {\cal Z}, n \neq 0, \; 2\leq j \leq D-1 \}$,

the transverse center of mass variables $\{ q^j , p^j , \; 2\leq j
\leq D-1 \}$,

$q^V$ and $p^U$.

\bigskip

Up to now we have considered a classical string.

At the quantum level one imposes the canonical commutation relations (CCR)
\begin{eqnarray}\label{ccr}
 [ \a^{i}_{n}, \a^{j}_{m} ] = n \; \delta_{n,-m}\, \delta^{i,j} \quad &,&
  \quad [ {\tilde \a^{i}_{n}, \a^{j}_{m} }] = n \; \delta_{n,-m}\,
\delta^{i,j}
	\; ,\cr\cr
 [ {\tilde \a^{i}_{n}}, \a^{j}_{m} ] &=& 0 \; ,\cr\cr
 [ q^i , p^j ] = i  \; \delta^{i,j} \quad &,&\quad [q^V, p^U ] = i
\end{eqnarray}
All other commutators being zero. An order prescription is needed to
unambiguously express the different physical operators in terms of
those obeying the CCR. The symmetric ordering is the simplest and more
convenient.

The space of string physical states is the the tensor product of the
Hilbert space of the $D-1$ center of mass variables $q_V, p_U, \{ q^j
, p^j , 2\leq j \leq D-1 \}$, times the Fock space of the harmonic
transverse modes. The string wave function is then the product of a
center of mass part times a harmonic oscillator part. The center of
mass can be taken, for example, as a plane wave. The harmonic oscillator part
can be written as the creation operators $ {\a^{j}_{n}}^{\dag},{\tilde{
{\a}^{j}_{n}}}^{\dag},\; n \geq 1,\;   2\leq j \leq D-1$ acting on the
oscillator ground state $|0>$.  This state is defined as usual by
$$
\a^{j}_{n}\; |0> = {\tilde \a^{j}_{n}}\;|0> = 0, \quad {\rm for~all~}
 n \geq 1,\;   2\leq j \leq D-1
$$
Notice that a string describes {\bf one particle}. The kind of
particle described depends on the oscillator wave function.
The mass and spin can take an infinite number of  different values.
That is, there is  an infinite number of different possibilities   for
the particle described by a string.

Let us consider the quantum mass spectrum. Upon symmetric ordering the
mass operator becomes,
\begin{equation}
m^2= {1 \o {2 \a'}}
\sum_{j=2}^{D-1}  \sum_{n=1}^{_\infty}
 \left[   \a^{j}_{n} \, \a^{j}_{-n} + \a^{j}_{-n} \, \a^{j}_{n}+
{\tilde  \a^{j}_{n}} \, {\tilde  \a^{j}_{-n}}+{\tilde  \a^{j}_{-n}} \,
{\tilde  \a^{j}_{n}} \right] \; .
\end{equation}
Using the commutation rules (\ref{ccr}) yields
\begin{equation}
m^2= {{D-2}\over  {\a'}}      \sum_{n=1}^{_\infty} n +
{1 \o {2 \a'}}\sum_{j=2}^{D-1}  \sum_{n=1}^{_\infty}
 \left[  {\a^{j}_{n}}^{\dag} \, \a^{j}_{n}+
{\tilde  {\a^{j}_{n}}}^{\dag} \,{\tilde  \a^{j}_{n}} \right]
\end{equation}
The divergent sum in the first term can be defined through analytic
continuation of the zeta function
\begin{equation}
\zeta(z) \equiv  \sum_{n=1}^{_\infty} { 1 \over {n^z}}
\end{equation}
One finds  $\zeta(-1) = -1/12$ \cite{grad}. Thus,
\begin{equation}\label{masaq}
m^2= -{{D-2}\over  {12 \a'}}
+ {1 \o {2 \a'}} \sum_{j=2}^{D-1}  \sum_{n=1}^{_\infty}
 \left[  {\a^{j}_{n}}^{\dag} \, \a^{j}_{n}+
{\tilde  {\a^{j}_{n}}}^{\dag} \,{\tilde  \a^{j}_{n}} \right]
\end{equation}
Hence, the string ground
state  $|0>$ has a negative mass squared
\begin{equation}
m_0^2 =  -{{D-2}\over  {12 \a'}}
\end{equation}
Such particles are called tachyons and exhibit unphysical
behaviours. When fermionic degrees of freedom are associated to the
string the ground state becomes massless (superstrings) \cite{gsw}.

Notice that the appearance of a negative mass square yields a
dispersion relation $E^2=p^2 - |m_0^2|$ similar to classical waves
when gravity (even newtonian) is taken into account
 (Jeans unstabilities) \cite{ll}.

Let us consider now excited states.

The constraints (\ref{vinm}) integrated on $\s$ from $0$ to $2\pi$
impose
\begin{equation}
\sum_{j=2}^{D-1}  \sum_{n=1}^{_\infty}
   (\a^{j}_{n})^{\dag} \, \a^{j}_{n} =
\sum_{j=2}^{D-1}  \sum_{n=1}^{_\infty}
{\tilde  {(\a^{j}_{n})}}^{\dag} \,{\tilde  \a^{j}_{n}}
\end{equation}
on the physical states. This means that the number of left and right
modes coincide in all physical states.

The first excited state is then described by
\begin{equation}\label{grav}
|i, j > = {\tilde  {(\a^{i}_{1})}}^{\dag} (\a^{j}_{1})^{\dag} \; |0>
\end{equation}
times the center of mass wave function. We see that this wavefunction
is a symmetric tensor in the space indices $i,j$. It describes
therefore a spin two particle plus a spin zero particle (the trace part).

{}From eqs.(\ref{masaq}-\ref{grav}) follows that
\begin{equation}
m^2\;|i, j > =  -{{D-26}\over  {12 \a'}} |i, j >
\end{equation}
This state is then a massless particle only for $D=26$. In such
critical dimension we have then a graviton (massless spin 2 particle)
and a dilaton  (massless spin 0 particle) as string modes of
excitation. For superstrings the critical dimension turns to be $D=10$
\cite{gsw}.

We shall consider, as usual, that only four space-time dimensions are
uncompactified. That is, we shall consider the strings as living
on the tensor  product of a curved four dimensional space-time with
lorentzian signature and a compact space which is there  to cancel
the anomalies. From now on strings will propagate in the
curved (physical) four dimensional space-time. However, we will find
instructive to study the case where this curved space-time has
dimensionality  $D$, where $D$ may be 2, 3, 4 or arbitrary.

\subsection{The string energy-momentum tensor and the string invariant size}

The spacetime string energy-momentum tensor follows (as usual) by taking the
functional derivative of the action (\ref{accion}) with respect to the
metric $ G_{AB} $ at the spacetime point $X$. This yields,

\begin{eqnarray}\label{tens}
\sqrt{-G}~ T^{AB}(X) = \frac{1}{2\pi \alpha'} \int d\sigma d\tau
\left( {\dot X}^A {\dot X}^B -X'^A X'^B \right)
\delta^{(D)}(X - X(\sigma, \tau) )
\end{eqnarray}
where dot and prime stands for $\partial/\partial\tau$ and
 $\partial/\partial\sigma $, respectively.

Notice that $X$ in eq.(\ref{tens}) is just a spacetime point whereas $
X(\sigma, \tau)$ stands for the string dynamical variables. One sees
from  the Dirac delta in eq.(\ref{tens}) that $ T^{AB}(X) $ vanishes
unless $X$ is exactly on the string world-sheet.
We shall not be interested in the detailed structure of the classical strings.
It is the more useful to integrate
the energy-momentum tensor (\ref{tens}) on a volume
that completely encloses the string. It takes then the  form \cite{ijm}
\begin{equation}
\Theta^{AB}(X^0)  =  \frac{1}{2\pi \alpha'} \int d\s d\tau
\left( {\dot X}^A {\dot X}^B -X'^A X'^B \right)
\delta(X^0 - X^0(\tau,\s ) ).
\end{equation}
When $ X^0 $ depends only on $\tau$, we can easily integrate over
$\tau$ with the result,
\begin{equation}\label{inten}
\Theta^{AB}(X^0)  =  \frac{1}{2\pi \alpha' |{\dot X}^0(\tau)|}
\int_0^{2\pi}d\s \left[ {\dot X}^A {\dot X}^B -X'^A X'^B \right]_
{\tau = \tau(X^0)}
\end{equation}

Another relevant physical magnitude for strings is the invariant
size. We define the invariant string size $ds^2$ using the  metric induced  on
the string world-sheet:
\begin{equation}\label{intinv}
ds^2 = G_{AB}(X) \, dX^A \, dX^B
\end{equation}
Inserting  $ dX^A = \partial_+X^A \; dx^+ +  \partial_-X^A \; dx^- $, into
eq.(\ref{intinv}) and taking into account the constraints
(\ref{conodos}) yields
\begin{equation}\label{tam}
 ds^2 = 2\, G_{AB}(X) \, \partial_+X^A \, \partial_-X^B
\left(d\tau^2 - d\s^2\right) =  G_{AB}(X) \, {\dot X}^A \, {\dot
X}^B \left(d\tau^2 - d\s^2\right) .
\end{equation}

Thus, we define the string size as the integral of
\begin{equation}\label{TC}
S(\s,\tau) \equiv \sqrt{G_{AB}(X)  \, {\dot X}^A \, {\dot X}^B}
\end{equation}
over $\s$ at fixed $\tau$. For a causal choice of the string time
$\tau$ we must have
\begin{equation}\label{sigtam}
G_{AB}(X)  \, {\dot X}^A \, {\dot X}^B \geq 0
\end{equation}
The equality sign here corresponds to a string behaving as
radiation. Such type of solutions always exist. For example any
$\sigma$-independent solution of eqs.(\ref{conouno}-\ref{conodos}).
Such solutions describe  massless geodesics.

Notice that the trace of the energy-momentum tensor eq.(\ref{tens})
is just the integral of $ S(\s,\tau)^2  $,
\begin{eqnarray}\label{traza}
\sqrt{-G}~ T_A^A(X) &=& \frac{1}{\pi \alpha'} \int d\sigma d\tau \; G_{AB}(X)
 {\dot X}^A {\dot X}^B\;
\delta^{(D)}(X - X(\sigma, \tau) ) \cr \cr
&=&  \frac{1}{\pi \alpha'} \int d\sigma d\tau \;
 S(\s,\tau)^2  \; \delta^{(D)}(X - X(\sigma, \tau) )
\end{eqnarray}
For strings behaving as radiation both $ T_A^A(X) $ and $ S $ vanish.

\subsection{Simple String Solutions in Minkowski Spacetime}

Let us now consider a circular string as a simple example of a string
solution in Minkowski spacetime.
\begin{eqnarray}\label{anilM}
X^0(\s,\tau) &=& \a' E \tau  \quad ,  \quad X^3(\s,\tau) =  \a' p \tau
\cr & & \cr
X^1(\s,\tau) &= &\a' m \cos\tau \cos\s = {{  \a' m} \over 2} \left[ \cos(\tau
+\s) +  \cos(\tau-\s) \right] \\ & & \cr
X^2(\s,\tau) &=& \a' m \cos\tau \sin\s = {{  \a' m} \over 2} \left[ \sin(\tau
+\s) -  \sin(\tau-\s) \right] \nonumber
\end{eqnarray}
This is obviously a solution of eqs.(\ref{dalam}) where
only the modes $n = \pm 1, j=1,2$ are excited.  The constraints
(\ref{vinm}) yields
$$
E^2 = p^2 + m^2
$$
Eqs.(\ref{anilM}) describe a circular string in the $X^1,X^2$ plane,
centered in the origin and
with an oscillating radius $\rho(\tau) =  \a' m \cos\tau $. In
addition  the string moves uniformly in the $z$-direction with speed
$p/E$. (That is, $p$ is its momentum in the $z$-direction).
The oscillation amplitude $m$ can be identified  with the string mass and $E$
with the string energy. Notice that the string time $\tau$ is here
proportional to the physical time $X^0$ [this solution is not in the
light-cone gauge (\ref{gauge})].

It is instructive to compute the  integrated energy-momentum tensor
(\ref{inten}) for this string solution. We find in the rest frame
($p=0$) that it takes the fluid form

\begin{eqnarray}
\Th_A^B = \left(\begin{array}{rrrr}
\rho & 0 & 0 & 0 \\
0  & -p & 0  & 0 \\
0 & 0  & -p & 0 \\
0  & 0  &  0 & 0 \end{array}\right)
\label{tani}
\end{eqnarray}
 where
\begin{equation}\label{eyp}
\rho = E = m \quad , \quad p = - {m \over 2} \cos(2\tau)
\end{equation}
We see that the total energy coincides with $m$ as one could expect
and that the (space averaged) pressure oscillates around zero. That is, the
string pressure goes through positive and negative values. The time
average of  $p$  on a period vanishes.  The string behaves then as
cold matter (massive particles).

The upper value of $p$ equals $E/2$. This is precisely the relation
between $E$ and $p$ for radiation (massless particles). (Notice that
this circular strings lives on a two-dimensional plane). The lower
value of $p$ correspond to the limiting value allowed by the strong
energy condition in General Relativity \cite{hawel}.
We shall see below that these two extreme values of $p$ appear for
strings in general cosmological spacetimes.

The invariant size of the string solution (\ref{anilM}) follows by
inserting eq.(\ref{anilM}) into eq.(\ref{tam}). We find
\begin{equation}
ds^2 = (\a' m)^2  \left(d\tau^2 - d\s^2\right)
\end{equation}
Therefore, this string solution has a constant size $2\pi \a' m$.

Another simple but instructive  solution in Minkowski spacetime
is a rotating straight string (a rotating rod) given by
\begin{eqnarray}\label{rota}
X^0(\s,\tau) &=& \a' m \tau  \quad ,
\cr & & \cr
X^1(\s,\tau) &= &\a' m \cos\tau \cos\s\; , \\ & & \cr
X^2(\s,\tau) &=& \a' m \sin\tau \cos\s \; ,
\nonumber
\end{eqnarray}
or in polar coordinates
$$
\rho = \a' m \; | \cos\s |  \quad ,  \quad \phi = \tau \; .
$$
That is, a straight string on the $X^1 - X^2 $ plane rotating around the
origin with an angular speed $ { 1 \o {  \a' m  }} $. Eqs. (\ref{rota})
identically fulfil the string equations and constraints
(\ref{dalam}-\ref{vinm}).

The energy momentum tensor for this rotating string takes the form:

\begin{eqnarray}
\Th_A^B = \left(\begin{array}{rrrr}
m & 0 & 0 & 0 \\
0  & {m \o 2}\cos(2\tau) & {m \o 2}\sin(2\tau)   & 0 \\
0 & {m \o 2}\sin(2\tau)    & - {m \o 2}\cos(2\tau) & 0 \\
0  & 0  &  0 & 0 \end{array}\right) \; .
\label{tanir}
\end{eqnarray}

This means an energy density $ \rho = m $. The spatial part of $ \Th_A^B $
becomes diagonal in a rotating frame and has $ \pm {m \o 2} $ as   eigenvalues.
That is, we again find  $ \pm {\rho \o 2} $ as extreme values for the pressure
in two space dimensional string solutions. The time average of the
stress tensor $\Th_i^j$ here vanishes indicating zero pressure (cold
matter behaviour).

One obtains the invariant size for a generic solution in Minkowski
spacetime inserting  the general solution (\ref{solM}) into
$\partial_+X^A \partial_-X_A$. This gives constant plus oscillatory
terms. In any case the  invariant string size is always {\bf bounded}
in Minkowski spacetimes. We shall see how differently behave strings in curved
spacetimes.

\section{\bf How to solve the string equations of motion in curved
spacetimes?}

There is no general method  to solve the string equations
of motion and constraints for arbitrary curved spacetime.

The so-called $\tau$-expansion method provides {\bf exact local} solutions
for any background. The basic idea goes as follows.
Suppose one is interested on the string behaviour near a given point of the
curved spacetime. Then, one chooses a conformal gauge such that $ \tau = 0 $
corresponds to such a point. For example, to study strings in cosmological
spacetimes near the initial singularity (cosmic time $ T = 0 $), one chooses
$ T(\s,\tau) $ such that  $ T(\s, 0 ) = 0 $. It is shown below that
this is indeed possible for generic string solutions. This expansion
was developped first in ref.\cite{sv,gsv} for inflationary universes.

Similarly, in order to study strings near the black hole singularity
 $ r = 0 $, one chooses $ r (\s,\tau) $ such that  $ r(\s, 0 ) = 0 $.

Once this gauge choice is done, the string equations of motion and constraints
can be solved in powers of $ \tau $. These powers may not be integer powers.
For example, one finds powers of $  \tau^{-{2\o{\a+1}}}  $ in FRW universes
with scale factor $ R(T) = c \; T^{\a} $. For Schwarzschild black
holes powers of
$ \tau^{2 \o 5} $ appear.

An approximate but general method is the expansion around center of mass
solutions \cite{dvs87,agn}. In this method one starts from an exact solution
of the geodesic equations
\begin{equation}\label{cm}
 \ddot q^{A}(\tau)  +   \Gamma^{A}_{BC} (q) \, \dot q^{B}(\tau)\,
 \dot q^{C}(\tau)  =  0
 \end{equation}
 The world-sheet time variable is here identified with the proper time of
  the center of mass trajectory. [Notice that eqs.(\ref{cm})  just
follow from the  string equations (\ref{conouno}) by dropping the
 $\s$-dependence].

Then one develops in perturbations around it. That is, one sets
 \begin{equation}\label{expan}
 X^A(\s,\tau)  =  q^A(\tau)  +  \eta^A(\s,\tau) +  \xi^A(\s,\tau)
+\ldots
\end{equation}
 Here $\eta^A(\s,\tau)$   obeys the linearized perturbation around
$q^A(\s,\tau)$ and  $ \xi^A(\s,\tau) $ the second order perturbation
equations \cite{dvs87}. These fluctuations obey coupled
ordinary differential equations that can be written systematically
 inserting eq.(\ref{expan}) into eqs.(\ref{conouno}-\ref{conodos}).
 [See ref.\cite{agn}  for more details].

Another  general approximation  method is the null string approach \cite{nic}.
In such approach the string equations of motion
and constraints are systematically
expanded in powers of $c$ (the speed of light in the world-sheet).
This corresponds to a small string tension expansion.
At zeroth order, the string is effectively equivalent to a continuous beam
of  massless particles labelled by the parameter $\sigma$. The points
on the string do not interact between them but they interact with the
gravitational background.

\medskip

For several spacetimes one can construct explicit string solutions
using specific properties of the background. This is the case of
singular plane waves, shock-waves, conical spacetimes and
the de Sitter universe. The string equations of motion and constraints
in the de Sitter spacetime are integrable in the inverse scattering
sense as shown in ref.\cite{prd}.
[The de Sitter universe is a symmetric space and
hence the string equations there correspond to a two dimensional integrable
sigma model].

\subsection{\bf The $\tau$-expansion}

Let us consider the intersection of the world-sheet with a singular or
non-singular point (or surface)
 of the spacetime like $ T(\s,\tau) = 0 $ or  $T(\s,\tau)
= T_o $ in a cosmological spacetime or $ r(\s,\tau) = 0 $ or  $
r(\s,\tau) = r_o $ in a Schwarzschild black-hole.

We can write the curve describing such intersection  with
the world-sheet as
\begin{equation}\label{hojhor}
x_+ = \chi(x_-),
\end{equation}
whenever this intersection is nondegenerate. (Here $ x_{\pm} \equiv
\tau \pm \s $).

Upon a conformal transformation,
\begin{equation}\label{confg}
x_+ \to x_+' = f(x_+) \quad , \quad x_- \to x_-' = g(x_-),
\end{equation}
we can map the curve (\ref{hojhor}) into $\tau' = 0$ by an appropriate
choice of $f$ and $g$ \cite{negro}. For example, we can choose
$$
 f(x_+) = x_+  \quad , \quad g(x_-) = -  \chi(x_-).
$$
This defines our choice of gauge. From now on, we rename $\tau'$ and $\s'$
by $\tau$ and $\s$, respectively. Notice that this choice does not
completely fix the gauge. We can still perform transformations that
leave the line $\tau = 0$ unchanged. This is the case for the
following class of conformal mappings
\begin{equation}\label{diago}
x_+ \to x_+' = \varphi(x_+) \quad , \quad x_- \to x_-' = -\varphi(-x_-),
\end{equation}
where $\varphi(x)$ is an arbitrary function. Eq. (\ref{diago}) can be
written as,
\begin{eqnarray}\label{diag}
\tau' &=& {1 \o 2} \left[ \varphi(\tau + \s) - \varphi(\s - \tau)
\right] = \tau \, \varphi'(\s)\, + {1 \o 6}\, \tau^3 \;
\varphi'''(\s)+ O(\tau^4) \cr \cr
\s' &=& {1 \o 2} \left[ \varphi(\tau + \s) + \varphi(\s - \tau)
\right] = \varphi(\s)\, + {1 \o 2}\, \tau^2 \;
\varphi''(\s)+ O(\tau^4)
\end{eqnarray}
The transformations  (\ref{diago}) represent a diagonal subgroup of
the set of left-right conformal transformations (\ref{confg}).

In summary, any  (non-degenerate) intersection of the world-sheet
with a spacetime  submanifold $T(\s,\tau) = T_o $ can be mapped into
$\tau = 0 $. This mapping is not unique, it is invariant under the
 diagonal conformal transformations.

Once this gauge has been imposed one studies the string equations of
motion and constraints (\ref{conouno}-\ref{conodos}) in powers of
$\tau$. The equations themselves determine the precise values of the
powers \cite{sv,gsv,negro}.

\subsection{\bf Global Solutions}

There is no general method to find solutions valid in the whole
world-sheet. However, many global solutions have been found in
physically relevant spacetimes.

First, there are spacetimes where the {\bf general} solution of the
string equations and constraints has been found. That is, shock-waves
\cite{ondch},
singular plane waves \cite{ondpl} and conical spacetimes  \cite{conico}.

Second, by making specific ansatz according to the symmetry of the
background, the string equations of motion can
be reduced to ordinary differential equations. Then, these   ordinary
differential equations can be solved globally by  numerical methods.
In this way, solutions valid in the whole worldsheet has been found in
cosmological spacetimes and black holes \cite{dls,din} -\cite{bhln}.

Third, the de Sitter spacetime can be treated by inverse scattering
methods. In this way exact string solutions has been constructed
systematically (see sec. VII and refs. \cite{dms} -\cite{igor}).

In all  cases where global solutions can be found,
the  $\tau$-expansion results are confirmed.

\section{\bf String propagation in cosmological spacetimes}

We obtain  in this section  physical string properties
from the  string solutions in cosmological spacetimes.

We consider strings in spatially homogeneous and isotropic
universes with metric

\begin{equation}\label{unicos}
ds^2 = (dT)^2 - R(T)^2 \sum_{i=1}^{D-1}(dX^i)^2 \; ,
\label{met}
\end{equation}

where $T$ is the cosmic time and the function $R(T)$
is called the scale factor. In terms of the conformal time

\begin{equation}
\eta = \int^{T} \frac{d T}{R(T)}~,
\label{tco}
\end{equation}

the metric (\ref{met}) takes the form

\begin{equation}
ds^2 = R(\eta)^2 \left[ (d\eta)^2 - \sum_{i=1}^{D-1}(dX^i)^2 \right] \; .
\label{metc}
\end{equation}

The classical string equations of motion can be written  here as
\begin{eqnarray}
& \displaystyle{\partial^2T
+ \; R(T)\; {{{\rm d}R}\over{\,{\rm d}T}}\;
\sum_{i=1}^{D-1}(\partial_\mu X^i)^2=0\; ,\label{eqmov}}\\
&\displaystyle{\partial_\mu
\left[R^2 \partial^\mu X^i\right]=0\,,
\qquad 1\leq i\leq D-1\,,\nonumber }
\end{eqnarray}
 and the constraints are

\begin{equation}
T_{\pm\pm}=(\partial_\pm T)^2 - R(T)^2 \; (\partial_\pm X^i)^2=0\,.
\label{vinc}
\end{equation}
The most relevants universes correspond to power type scale factors.
That is,
\begin{equation}\label{factor}
 R(T) = a \; T^{\a} = A \; \eta^{k/2} \; ,
\end{equation}
where $ \a = {k \o {k+2}}$ and $ k = {{2 \a} \o {1 - \a}} $.

For different values of the exponents we
have either FRW or inflationary universes.
$$
{\rm FRW:~} 0<k \le \infty, \; 0<\a\le 1  = \cases{
	\a = 1/2 , \; k=2, & {\rm radiation dominated}, \cr
	\a = 2/3, \; k=4, & {\rm matter dominated},	\cr
	 \a = 1 , \; k = \infty, &  {\rm `stringy'}. \cr}
$$

$$
{\rm Inflationary:}\;  -\infty <k < 0 , \; \a< 0  {\rm ~and~}\a>1
= \cases{
\a =  \infty , \; k=-2, &  $R(T) = e^{HT} , \;$ {\rm de ~ Sitter}, \cr
	\a > 1, \; k<-2 ,  & {\rm power inflation}, \cr
 \a < 0 , \; -2 < k < 0 , & {\rm superinflationary}. \cr}
$$

The denomination  `stringy' comes from the fact that such backgrounds
follow as solution of the string effective equations \cite{tse}.
Inflationary universes are those with accelerated expansion. That is,
$$
{{d^2R(T)}\over {dT^2}} > 0 \; .
$$

The string equations of motion and constraints take then the following
form using the conformal time $\eta$:
\begin{eqnarray}\label{eqeta}
& \displaystyle{
{\ddot \eta} - \eta'' + {k \o {2 \eta}}\; \left\{{\dot \eta}^2 -
\eta'^2 +\sum_{i=1}^{D-1} \left[ ({\dot X}^i)^2 - (X'^i)^2 \right]
\right\}=0} & \; , \cr
& \displaystyle{
{\ddot X}^i - X''^{i} + {k \o {\eta}}\; \left[{\dot \eta}{\dot X}^i -
\eta' X'^i \right]=0} \; , \quad 1 \leq i \leq D-1 &  \; ,\\
&  \displaystyle{
({\dot \eta}\pm \eta')^2 - \sum_{i=1}^{D-1} ({\dot X}^i\pm X'^i
)^2}=0  \; , \nonumber
\end{eqnarray}

where prime   and dot  stand for $\partial_{\s}$ and  $\partial_{\tau}$,
respectively.

\medskip

As we will see below, once appropriately averaged $\Theta^{AB}(X)$ takes
the fluid form for string solutions in cosmological spacetimes,
allowing us to define the string pressure $p$ and energy
density $\rho$ :

\begin{eqnarray}\label{tmnflu}
<\Theta_A^B> = \left(\begin{array}{rrrr}
\rho & 0 & \cdots & 0 \\
0  & -p & \cdots & 0 \\
\cdots & \cdots  & \cdots & 0 \\
0  & 0  &  \cdots & -p \end{array}\right) \; .
\label{tflu}
\end{eqnarray}

Notice that the continuity equation

\begin{eqnarray}
D^A\;\Theta_A^B = 0  \; ,\nonumber
\end{eqnarray}

takes here the form

\begin{equation}
 {{d\rho}\over{dT}} + (D-1)\, H\, (p + \rho ) = 0
\label{cont}
\end{equation}

where $ H \equiv {{1}\over {R}} {{dR}\over{dT}} $.

For an equation of state of the type of a perfect fluid, that is

\begin{equation}
p = (\gamma - 1 )\;\rho \qquad , \qquad \gamma = {\rm~constant},
\label{eflu}
\end{equation}

 eqs.(\ref{cont}) and (\ref{eflu}) can be easily integrated with the
result

\begin{equation}
\rho = \rho_0~R^{\,\gamma(1-D)}~~.
\label{ror}
\end{equation}

For $\gamma = 1$ this corresponds to cold matter $(p = 0)$ and
for $\gamma = \frac{D}{D-1} $ this describes radiation with
$p = {{ \rho}\over{D-1}}$.

The speed of sound in a fluid is given by
$$
v_s = \sqrt{\left({{\partial p}\over{\partial  \rho}}\right)|_s} =
\sqrt{\gamma - 1}
$$
This relation makes sense for $ 1 \geq \gamma - 1 \geq 0 $. For $
\gamma - 1 < 0 $, there are no sound waves since the perturbations in
$ p $ and $  \rho $ obey elliptic evolution equations. For  $
\gamma - 1 > 1 $ causality would be violated. Actually, for strings we
 always find $   - \frac{1}{D-1}\leq \gamma - 1 \leq \frac{1}{D-1}
$.

\subsection{Strings in cosmological universes: the $\tau$-expansion at
work}

Let us consider strings in inflationary universes with scale factor
(\ref{factor}) and $ k < 0 $. In order to apply the  $\tau$-expansion
we fix the gauge such that
\begin{equation}\label{etacero}
\eta(\tau = 0, \s) = 0 \; .
\end{equation}
As explained above, this is always possible for generic string
solutions and it leaves still the freedom of the transformations
 (\ref{diago}). Notice that $\eta \to 0 $ corresponds in the
inflationary case to large scale factors $ R \to \infty $.

The behaviour of $\eta(\tau, \s)$ and $X^i(\tau, \s)$ for $\tau \to 0$
follows from eq.(\ref{eqeta}) where we use  eq.(\ref{etacero}) and
assume that $X^i(\tau , \s)$ is regular at $\tau = 0$. One finds
\cite{gsv}
\begin{eqnarray}
\eta(\tau, \s) &\buildrel{\tau \to 0}\over=&  \eta_o(\s)\; \tau \;
\left[ 1 + O(\tau^2) \right] +\l_o(\s)\; \tau^{2-k} \;
\left[ 1 + O(\tau^2) \right] \cr
&+& \zeta_o(\s)\; \tau^{1-2k} \;
\left[ 1 + O(\tau^2) \right] + O(\tau^{1-3k}) \; ,\cr  & & \quad
\quad \\
X^i(\tau , \s) &\buildrel{\tau \to 0}\over=& A^i( \s)\left[ 1 +
O(\tau^2) \right] + \tau^{1-k} \; B^i( \s)\;
\left[ 1 + O(\tau^2, \tau^{1-k}  ) \right]  \; ,\cr
& & \quad  \quad 1\leq i \leq D-1 \;. \nonumber
\end{eqnarray}

The solutions appear as a  series in  powers of $\tau^2$ and $
\tau^{1-k} $. In the special case where $k$ is rational, say $k = -{l
\over n}\; , \; l,n= $integers, logarithmic terms in $\tau$ appear in
addition. This happens, for example in de Sitter spacetime ($ k = - 2
$ ) and in Minkowski spacetime ($ k = 0 $).

The coefficients in  eq.(\ref{inflt0}) result related as follows
\begin{eqnarray}
 \sum_{i=1}^{D-1}\, B^i (\s)\; A'^{i}( \s) &=& 0 \; , \;
  \eta_o(\s) = \sqrt{ \sum_{i=1}^{D-1}\,\left[A'^{i}( \s)\right]^2}\;
\cr
\l_o(\s) &=& \displaystyle{
{{2k}\over{(2-k)(1-k)}}\; {{ \sum_{i=1}^{D-1}\, B'^{i}( \s)\;
A'^{i}( \s) }\o { \eta_o(\s)}}} \; .
\end{eqnarray}
Moreover, one can use the residual conformal invariance  (\ref{diago})
to set $  \eta_o(\s) \equiv 1 $. One finally obtains for inflationary
universes \cite{gsv},
\begin{eqnarray}\label{inflt0}
\eta(\tau, \s) &\buildrel{\tau \to 0}\over=&   \tau \;
\left[ 1 + O(\tau^2) \right] +\l_o(\s)\; \tau^{2-k} \;
\left[ 1 + O(\tau^2) \right] \cr
&+& \zeta_o(\s)\; \tau^{1-2k} \;
\left[ 1 + O(\tau^2) \right] + O(\tau^{1-3k}) \; ,\cr  & & \quad
\quad \\
X^i(\tau , \s) &\buildrel{\tau \to 0}\over=& A^i( \s)\left[ 1 +
O(\tau^2) \right] + \tau^{1-k} \; B^i( \s)\;
\left[ 1 + O(\tau^2, \tau^{1-k}  ) \right] \; ,\cr
& & \quad  \quad 1\leq i \leq D-1 \;. \nonumber
\end{eqnarray}
where
\begin{eqnarray}\label{relinf}
 & & \sum_{i=1}^{D-1}  B^i( \s)\; A'^{i}( \s) = 0  \quad , \quad
  \sum_{i=1}^{D-1}\,\left[A'^{i}( \s)\right]^2  = 1 \cr
\l_o(\s) &=& {{2k}\over{(2-k)(1-k)}}\; \sum_{i=1}^{D-1}\, B'^{i}( \s)\;
A'^{i}( \s) \quad , \quad
\zeta_o (\s) = {{(1-k)^2}\over {2(1-2k)}}\;
\sum_{i=1}^{D-1}\,\left[B'^{i}( \s)\right]^2   \; .
\end{eqnarray}
Here $ A^{i}( \s), \; B^{i}( \s), \;  1\leq i \leq D-1 $ are the
initial $ (\tau = 0 ) $ string coordinates and momenta.

We see that the solution depends on $2(D-2)$ independent functions
among the
$ A^{i}( \s) $ and $ B^i( \s) \; , \; 1 \leq i \leq D-1 $. All
coefficients (including the higher orders not written in
eq.(\ref{inflt0})) express in terms of the $ A^{i}( \s) $ and $ B^i(
\s) $. Therefore, the counting of degrees of freedom turns to be the
same as in Minkowski spacetime: only the  $2(D-2)$  {\bf transverse}
coordinates are physical.

It must be noticed that
$$
{\dot  X}^i(\tau , \s) \buildrel{\tau \to 0}\over= (1-k)  \; \tau^{-k}
\; B^i( \s)\; \left[ 1 + O(\tau^2, \tau^{1-k}  ) \right] \to 0
$$
$$
X'^{i}(\tau , \s) \buildrel{\tau \to 0}\over= A'^{i}( \s) \neq 0
$$
That is, $ \partial_{\s}X^i$ is {\bf larger} than  $
\partial_{\tau}X^i$  for $ R \to \infty $. This is the opposite to a
point particle behaviour. For a point particle,  $ \partial_{\s}X^i
\equiv 0 $ and $\partial_{\tau}X^i = p^i \neq 0 $.

\bigskip

Let us now apply the $\tau$-expansion to strings in FRW
universes. That is $ k > 0 $ in the scale factor (\ref{factor}).

We fix again the gauge according to eq.(\ref{etacero}). It must be
noticed that now $\eta \to 0 $ corresponds to $R \to 0 $ since $ k > 0
$. That is the  $\tau$-expansion applies near the singularity (big
bang) of the spacetime.

After  calculations analogous to the inflationary case, one finds
from eqs.(\ref{eqeta}) for FRW universes $(k>0)$ \cite{gsv}
\begin{eqnarray}\label{frw0}
\eta(\tau, \s) &\buildrel{\tau \to 0}\over=&   \tau^{\frac{1}{k+1}} \;
\left[ 1 + O(\tau^2) \right] +\eta_1( \s)\;  \tau^{2-\frac{1}{k+1}}
\left[ 1 + O(\tau^2) \right] \cr & & \quad  \quad \\
X^i(\tau , \s) &\buildrel{\tau \to 0}\over=& A^i( \s)\left[ 1 +
O(\tau^2) \right] + \tau^{\frac{1}{k+1}}\; B^i( \s)\;
\left[ 1 + O(\tau^2 ) \right] + \tau^{2-\frac{1}{k+1}}\; C^i( \s)\;
\left[ 1 + O(\tau^2 ) \right]  \; ,\cr
& & \quad  \quad 1\leq i \leq D-1 \;. \nonumber
\end{eqnarray}
where the residual conformal invariance  (\ref{diago}) has been used.

The string equations of motion and constraints impose the following relations
on the coefficients:
\begin{eqnarray}\label{relfrw}
 \sum_{i=1}^{D-1}  \left[ B^i( \s)\right]^2 &=&1\,  \quad , \quad
   \sum_{i=1}^{D-1}\, B^i( \s)\; A'^{i}( \s) = 0 \; , \cr
 C^i( \s) &=& - \eta_1( \s) \;  B^i( \s)  \quad , \quad
\eta_1( \s) ={{(k+1)^2}\o{4(2k+1)}} \;
\sum_{i=1}^{D-1}\,\left[A'^{i}( \s) \right]^2   \; .
\end{eqnarray}
$ A^{i}( \s), \; B^{i}( \s), \;  1\leq i \leq D-1 $ turn to be  the
initial $ (\tau = 0 ) $ string coordinates and momenta.

We again find that the string solution is determined by $2(D-2)$
independent functions indicating that only the transverse coordinates
are physical. As we see, the   $\tau$-expansion produces for FRW universes
the string solution
as a  series in  powers of $\tau^2$ and $ \tau^{\frac{2k}{1+k}} $.

For large $R$ the spacetime curvature tends to zero as $T^{-2}\simeq
R^{-2/\alpha}$ in
FRW spacetimes. That is, for $ T \to \infty $. In order to analyze
the string behaviour in such regime it is convenient to choose the gauge
\begin{equation}
\eta = \eta(\tau) \to \infty \quad ,  \quad \mbox{for}~ \tau\to \infty
\end{equation}
[This is a slight generalization of the previous gauge choices at finite
$\tau$].

We then find from eq.(\ref{eqeta})
\begin{eqnarray}\label{frwinf}
\eta(\tau, \s) &\buildrel{\tau \to \infty}\over=&   \tau^{\frac{2}{k+2}} \;
\to \infty \; , \cr & & \quad  \quad \\
X^i(\tau , \s) &\buildrel{\tau \to \infty}\over=& \frac{1}{k+2}\;
 \tau^{-\frac{k}{k+2}}\; \left[ f^+_i(\s + \tau) +  f^-_i(\s - \tau) \right]
  \; ,\cr & & \quad  \quad 1\leq i \leq D-1 \;. \nonumber
\end{eqnarray}
where the $f^{\pm}_i(x)$ are arbitrary periodic functions of $x$
$$
f^{\pm}_i(x+ 2 \pi)=f^{\pm}_i(x) \; ,
$$
obeying the pair of constraints:
\begin{equation}
 \sum_{i=1}^{D-1} \left[ f'^{\pm}_i(x) \right]^2 = 1\; .
\end{equation}
We obtain from eq.(\ref{frwinf})
\begin{equation}
T(\tau ) \buildrel{\tau \to \infty}\over= {{2 \sqrt{A} } \o {k
+ 2 }}  \; \tau\to \infty \quad \mbox{and}   \quad R \buildrel{\tau \to
\infty}\over=  \tau^{\frac{k}{k+2}}\; \to \infty\; .
\end{equation}
In short,  the string solutions in this regime are asymptotically
Minkowski solutions (\ref{solM}) {\em scaled} by a factor $ R^{-1}
$. This is not unexpected since the spacetime curvature vanishes for
this regime. The counting of degrees of freedom is again as in
Minkowski spacetime.

\bigskip

We have determined the string behaviour for $ R \to \infty $ in
inflationary and FRW universes and for $ R \to 0 $ in FRW
universes. Let us now compute  for such regimes the string physical properties,
energy-momentum and size.

The calculation of $ \Theta^{AB}(T)$ in the different limiting regimes
follows from eq.(\ref{inten}) since  $\eta = \eta(\tau) $
asymptotically (both for $\tau \to 0 $ and $ \tau \to \infty$).

\medskip

Let us start by considering the inflationary universes for $ R  \to
\infty$. We find for the (integrated) energy-momentum tensor
from eqs.(\ref{inten}) and (\ref{inflt0})

\begin{eqnarray}\label{tenmof}
\rho(T) &=& \Theta^{00}(T)  \buildrel{\tau \to 0}\over= {R\over
{\alpha'}} \to +\infty \; , \cr
& & \quad  \quad \\
 \Theta^i_j(T)  &\buildrel{\tau \to 0}\over=& {R\over{\alpha'}}\;
\int_0^{2\pi} {{d\s}\over
{2\pi}}\; A'^{i}( \s)\; A'^{j}( \s) \; \to \infty  \; , \cr
& & \quad  \quad \\
 \Theta^0_i(T) &\buildrel{\tau \to 0}\over=&  - {1\over{\alpha'}}\;
\int_0^{2\pi} {{d\s}\over {2\pi}}\; B^{i}( \s) \nonumber
\end{eqnarray}
where we also used the spacetime metric $G_{00} = 1 , \; G_{ij} = -
R^2 \; \delta_{ij} $.

Recall that $ \displaystyle{\sum_{i=1}^{D-1}  [A'^{i}( \s)]^2 = 1} $.

We see that the energy density $\rho(T)$ diverges for $  R \to \infty
$. The stress tensor $[ \Theta^i_j(T) ]$ is not diagonal but it is
given by a {\bf positive definite} matrix. Such matrix has then
positive eigenvalues and therefore tells us that the pressure is {\bf
negative} [compare with eq.(\ref{tmnflu})].
This is the {\bf unstable} string behaviour.
That is, the energy tends to $+  \infty$ and  the pressure to  $-  \infty$.
At the same time the energy flux density $ \Theta^0_i(T) $ stands bounded.

The string size $S$ in cosmological spacetimes (\ref{unicos}) takes the  form
\begin{equation}\label{tamco}
 S^2 = G_{AB}(X) \, {\dot X}^A \, {\dot X}^B = {\dot T}^2
- R^2\sum_{i=1}^{D-1} ({\dot X}^i)^2
\end{equation}
For $\tau \to 0 , \; R\to \infty $ in inflationary spacetimes, we find
using   eq.(\ref{inflt0}) that the first term dominates in eq.(\ref{tamco})
\begin{equation}\label{Taminf}
S \buildrel{\tau \to 0}\over= c\; \tau^{k/2} \simeq R \to \infty \; ,
\end{equation}
where $c$ is a constant. We see that the string grows infinitely big when the
universe inflates. The string size being proportional to the scale factor
and also to the string energy [$\rho$ in eq.(\ref{tenmof})].
These explosive  growings characterize the string unstable behaviour.

\medskip

Let us now consider FRW universes ($k>0$) for $ \tau \to 0,\;  R \to 0 $.
The (integrated) energy-momentum tensor in such regime takes the form

\begin{eqnarray}\label{tenmoFRW}
\rho(T) &=& \Theta^{00}(T)  \buildrel{\tau \to 0}\over= {1\over
{\alpha'(k+1)\; R}} \to +\infty \; , \cr
& & \quad  \quad \\
 \Theta^i_j(T)  &\buildrel{\tau \to 0}\over=& - {1\over{\alpha'(k+1)\; R}}
\;\int_0^{2\pi} {{d\s}\over
{2\pi}}\; B^{i}( \s)\; B^{j}( \s) \; \to \infty  \; , \cr
& & \quad  \quad \\
 \Theta^0_i(T) &\buildrel{\tau \to 0}\over=&  {1\over{\alpha'(k+1)\; R}}
\;\int_0^{2\pi} {{d\s}\over {2\pi}}\; B^{i}( \s) \nonumber
\end{eqnarray}

where we used  eqs.(\ref{inten}) and (\ref{frw0}). [Recall that
$\sum_{i=1}^{D-1}\, \left[ B^i( \s)\right]^2 = 1 $ ].

We see in eq.(\ref{tenmoFRW}) that
the stress tensor $( \Theta^i_j(T) )$ is not diagonal but it is
given by a {\bf negative definite} matrix. Such matrix has then
 negative eigenvalues and therefore tells us that the pressure is {\bf
positive}. This string behaviour is dual to the previous unstable behaviour.

We find from eq.(\ref{tamco}) for the string size $S$ in FRW universes
for $R\to 0$,
\begin{equation}\label{tamfrw}
S \buildrel{\tau \to 0}\over= \sqrt{\sum_{i=1}^{D-1}  [A'^{i}( \s)]^2} \;
\tau^{{k \o {2(k+1)}}} \simeq  \sqrt{\sum_{i=1}^{D-1}  [A'^{i}( \s)]^2}
\; R \to 0
\end{equation}

In this dual to  unstable behaviour, the string size {\bf
vanishes}. That is, the string
starts at the big bang with zero size.

In summary,  the energy tends to $+  \infty$,  the pressure also to
$+  \infty$
and the size tends to zero in the  {\bf dual to  unstable} behaviour.

In this regime strings  behave as radiation (massless particles). Recall that
the string size is proportional to the trace of the energy momentum tensor
[see eq.(\ref{traza})].

Let us finally  consider FRW universes ($k>0$) for $ \tau \to \infty ,\;
R \to  \infty $. There, the (integrated) energy-momentum tensor takes the form

\begin{eqnarray}\label{tenmoFRW2}
\rho(T) &=& \Theta^{00}(T) \buildrel{\tau \to \infty }\over= {{2 \sqrt A }\over
{\alpha'(k+2)\; }} \; , \cr
& & \quad  \quad \\
 \Theta^i_j(T)  &\buildrel{\tau \to \infty }\over=&  {R\over{\alpha'}}
\;\int_0^{2\pi} {{d\s}\over
{2\pi}}\;  \left[ f'^+_i \;  f'^-_j +  f'^-_i \;  f'^+_j \right] \; , \cr
& & \quad  \quad \\
 \Theta^0_i(T) &\buildrel{\tau \to \infty }\over=&  0 \; . \nonumber
\end{eqnarray}

where we used  eqs.(\ref{inten}) and (\ref{frwinf}). The string energy
here tends to a bounded constant. Since the $ f'^{\pm}_i(\s \pm \tau)$ are
periodic functions, their average on a period of time vanishes:
$$
\int_0^{2\pi} d\tau d\s \;  f'^+_i(\s + \tau)\;  f'^-_i(\s - \tau) =
\frac{1}{2}\int_{-2\pi}^{+2\pi} dx_-  \; \int_{|x_-|}^{4\pi-|x_-|} dx_+ \;
 f'^+_i(x_+)   \;  f'^-_i(x_-) = 0\; .
$$
Hence, the pressure vanishes when averaged over a string oscillation.
This is the {\bf stable} string behaviour. Here strings behave as dust
(cold matter) with $ p = 0 $ as equation of state.

 The string size $S$ follows eq.(\ref{tamco}) and  eq.(\ref{frwinf}),
\begin{equation}
S^2 \buildrel{\tau \to \infty}\over= {2 \o {(k+2)^2}}\;   \left[ 1 +
\sum_{i=1}^{D-1} f'^+_i f'^-_i \right] \;
\end{equation}
The string size is thus bounded for $R \to \infty$. Moreover, averaging
over  a period of time, we find
$$
{\bar S} = {\sqrt2 \o {k+2}} \; .
$$

\subsection{The perfect gas of strings}

Our aim is to provide a string description of matter appropriate to
the early universe.

Let us consider classical strings interacting with the cosmological
spacetime background and neglect their mutual interactions. That is,
we consider a  perfect gas of strings under the cosmic gravitational field.
The energy-momentum of such gas is just the sum over individual string
solutions. For each string the results of section IV.A apply.

We assume arbitrary initial data for the strings. Therefore, summing
over solutions is equivalent to {\em average} over the initial data
$ A^{i}( \s), \; B^{i}( \s), \;  1\leq i \leq D-1 $.

\medskip

For inflationary spacetimes the relevant quantity to average is
the integral
$$
\int_0^{2\pi} {{d\s}\over {2\pi}}\; A'^{i}( \s)\; A'^{j}( \s)
$$
that appears in the energy-momentum tensor eq.(\ref{tenmof}).

The average  will vanish for $ i \neq j $ and will be $ i$-independent
for $ i = j $ because
 we treat independently  the different $ A^{i}( \s) $.
Taking into account the constraint (\ref{relinf})
$$
 \sum_{i=1}^{D-1}\,\left[A'^{i}( \s) \right]^2 = 1 \; ,
$$
finally yields,
\begin{equation}
<\int_0^{2\pi} {{d\s}\over {2\pi}}\; A'^{i}( \s)\; A'^{j}( \s)> =
{{\delta_{ij}}\o {D-1}} \; .
\end{equation}
Therefore the stress tensor (\ref{tenmof}) takes for  unstable strings
the fluid form for  $R \to \infty$,
$$
 <\Theta^i_j(T)>  \buildrel{R \to \infty}\over= - {p \o {D-1}}\;
\delta_{ij} \; ,
$$
with
\begin{equation}\label{estin}
 p \equiv -{R\over{\alpha' (D-1)}}\; = - { {\rho}\o  {D-1}}\to -\infty \; ,
\end{equation}
where we used the expression for $\rho$ in eq.(\ref{tenmof}). Recall
that the string size also grows as $R$ for $R\to\infty$ [eq.(\ref{Taminf})].

 This equation of state exactly saturates the strong
energy condition in general relativity.

The unstable string behaviour corresponds to the critical case of the
so-called coasting universe \cite{ell,tur}. In other words,
the perfect gas of
strings provide a {\it concrete} matter realization of such cosmological
model. Till now, no form of matter was known to describe coasting universes
\cite{ell}.

\bigskip

The quantity to average in FRW spacetimes for small $ R $ is
$$
\int_0^{2\pi} {{d\s}\over {2\pi}}\; B^{i}( \s)\; B^{j}( \s)
$$
which appears in the energy-momentum tensor eq.(\ref{tenmoFRW}).

Following the same argument as above for the average over $A^{i}( \s)$
and recalling that  (\ref{relfrw})
$$
 \sum_{i=1}^{D-1}\,\left[B^{i}( \s) \right]^2 = 1 \; ,
$$
we find
\begin{equation}
<\int_0^{2\pi} {{d\s}\over {2\pi}}\; B^{i}( \s)\; B^{j}( \s)> =
{{\delta_{ij}}\o {D-1}}
\end{equation}
Hence the string energy-momentum tensor (\ref{tenmoFRW}) for dual to
unstable strings takes the fluid form for $R \to 0$,
$$
 <\Theta^i_j(T)>  \buildrel{R \to 0}\over= - {p \o {D-1}}\; \delta_{ij}
$$
with
$$
p \equiv {1\over{\alpha' (D-1)(k+1) \; R}}\; \buildrel{R \to 0}\over=
+ { {\rho}\o  {D-1}}\to +\infty
$$
where we used the expression for $\rho$ in eq.(\ref{tenmoFRW}). Recall
that the string size vanishes as $R$, as $R$ vanishes [eq.(\ref{tamfrw})].

Therefore  in the dual to unstable case, strings behave as radiation
(massless particles).

\bigskip

For large $R$ in FRW spacetimes we must average independently over the
functions $f^+_i$ and  $f^-_j, \; 1 \leq i, j \leq D-1$. We find then
from eqs.(\ref{tenmoFRW2})
$$
 <\Theta^i_j(T)> \buildrel{R \to \infty}\over = 0 \; .
$$
That is,  the equation of state
$$
p=0 \; .
$$
The string energy and size are bounded in this  regime.
The strings behave for the stable regime as dust (cold matter). That
is, they  behave as massive particles.

\bigskip

In conclusion, an ideal gas of classical strings in cosmological universes
exhibit three different thermodynamical behaviours,
all of perfect fluid type:

\medskip

\begin{itemize}

\item (1) For  inflationary universes and $R\to\infty$ unstable strings:
negative pressure gas with $ p= - { {\rho}\o  {D-1}} $.

\item (2) Dual behaviour in FRW  universes and $R\to 0$: positive pressure
gas similar to radiation,
$ p= + { {\rho}\o  {D-1}} $.

\item (3) Stable strings in FRW  universes and $R\to \infty$: positive
pressure gas similar to cold  matter, $ p = 0 $.

\end{itemize}

\medskip

Tables 1 and 2 summarize the main string properties for any scale
factor $ R(T) $.

\vskip 30pt

\begin{centerline}
 {TABLE 1. String energy and pressure as obtained from exact}
 {string solutions for various expansion factors  $R(T)$.}

\bigskip

\bigskip

{\bf STRING PROPERTIES FOR ARBITRARY $R(T)$}

\end{centerline}

\vskip 20pt
\begin{tabular}{|l|l|l|l|}\hline
D-Dimensional spacetimes:&  $ $ & $ $ &  Equation  of State: \\
three asymptotic& $~~$ Energy  & $~~~~$Pressure  & $ $   \\
behaviours (u, d, s)&  $ $ & $ $ & $~~~$\\ \hline
$ $&  $ $ & $ $ & $ $\\
(i) unstable for $R\to\infty$ & $E_u  \buildrel{R \to \infty}\over = u
\; R \to \infty$ & $P_u = -{E_u\over{D-1}}\to-\infty$ &
`stringy' \\
$ $&  $ $ & $ $ & $ $\\
(ii) dual to (i) for $R \to 0$ & $E_d \buildrel{R \to 0}\over = d /R
\to \infty$ & $P_d = + {E_d\over{D-1}} \to \infty$ & radiation \\
$ $&  $ $ & $ $ & $ $\\
(iii) stable for  $R\to\infty$ & $E_s = \;$ constant & $P_s = 0 $ &
 dust (cold matter) \\
$ $&  $ $ & $ $ & $ $\\ \hline
\end{tabular}

\bigskip

\vskip 20pt

\begin{centerline}
 {TABLE 2. The string energy density and pressure}
 {for a gas of strings can be summarized by the formulas}

{below which become exact for $R \to 0$ and for $R \to \infty$.}

\vskip 20pt

\bigskip

{\bf STRING ENERGY DENSITY AND PRESSURE}

{\bf  FOR ARBITRARY $R(T)$}

\end{centerline}
\vskip 20pt
\begin{tabular}{|l|l|l|}\hline
$ $& Energy density: $~\rho \equiv E/R^{D-1}$ & \hspace{10mm}Pressure  \\
$ $&  $ $ & $ $ \\ \hline

Qualitatively correct &  $ $ & $ $\\
formulas for all R and D &
$ ~~\rho = \left( u \; R + {{d} \over R} + s \right) {1 \over
{R^{D-1}}} $ &
$ p  = {1 \over {D-1}} \left( {d \over R} -  u \; R\right) {1 \over
{R^{D-1}}} $ \\
$ $&  $ $ & $ $ \\ \hline
\end{tabular}

\bigskip

\newpage
Finally, notice that strings continuously evolve from one type of
behaviour to the other two. This is explicitly seen from the string
solutions in refs.\cite{dms} - \cite{dls} . For example the string
described by $q_-(\sigma, \tau)$ for $ \tau > 0$ shows unstable
behaviour for $ \tau \to 0 $, dual behaviour for $ \tau \to \tau_0 =
1.246450...$ and stable behaviour for $ \tau \to \infty $ .

\vskip 30pt

\begin{centerline}
 {TABLE 3. The {\bf self-consistent} cosmological solution}
{ of the Einstein equations in General Relativity}

{ with the string gas as source.}

\bigskip

{\bf STRING COSMOLOGY IN GENERAL RELATIVITY}

\end{centerline}
\vskip 20pt
\begin{tabular}{|l|l|l|}\hline
Einstein equations & \hspace{4mm}Expansion factor  &\hspace{6mm} Temperature \\
(no dilaton field) & \hspace{12mm} $R(T)$ &  \hspace{13mm}$T(R)$ \\
\hline
$ $&  $ $ & $ $ \\
$ T \to 0 $ & ${D \over 2}
\left[{{2 d}\over {(D-1)(D-2)}}\right]^{1 \over D}~ T^{2 \over D}$&
\hspace{4mm}${{dD}\over{S(D-1)}}\;1/R$\\
$ $&  $ $ & $ $ \\ \hline
$ $&  $ $ & $ $ \\
$ T \to \infty$ &$\left[{{(D-1) s}\over {2(D-2)}}\right]^{1 \over
{D-1}}~ T^{2 \over {D-1}}$&usual matter \\
$ $ &  $ $ &  dominated behaviour  \\ \hline
\end{tabular}

\vskip 40pt

\section{\bf Self-consistent string cosmology}

In the previous section we investigated the propagation of test
strings in cosmological spacetimes. Let us now investigate how
the Einstein equations in General Relativity and
the effective equations of string theory (beta functions) can be verified
{\bf self-consistently} with our string solutions as sources.

We shall assume a gas of classical strings neglecting interactions as
string splitting and coalescing. We will look for cosmological
solutions described by metrics of the type (\ref{met}). It is natural
to assume that the background will have the same symmetry as the sources.
That is, we assume that the string gas is homogeneous, described by a
density energy $ \rho = \rho(T) $ and a pressure $ p = p(T) $.
In  the  effective equations of string theory we consider a space
independent dilaton field. Antisymmetric tensor fields wil be ignored.

\subsection{\bf String Dominated Universes in General Relativity
(no dilaton field)}

The Einstein equations for the geometry (\ref{met}) take the form

\begin{eqnarray}
{1 \over 2}~(D-1)(D-2)~ H^2 & = & \rho \quad, \nonumber \\
(D-2){\dot H} + p + \rho & = & 0 \quad .
\label{eins}
\end{eqnarray}

where $ H \equiv {{dR}\over{dT}}/ R $. We know $p$ and $\rho$ as
functions of $R$ in asymptotic cases. For large $ R $, the unstable
strings dominate [eq.(\ref{estin})] and we have for inflationary spacetimes
\begin{equation}
\rho  = u \; R^{2-D} ~~~,~~~p = -{{\rho} \over {D-1}}\;
\quad {\rm  for ~}R \to \infty
\end{equation}

For small $ R $, the dual regime dominates with

\begin{equation}
\rho  = d \; R^{-D} ~~~,~~~p = +{ {\rho}\over {D-1}}\;\quad {\rm
for ~} R\to 0
\end{equation}

We also know that stable solutions may be present with a contribution
$ \simeq R^{1-D} $ to $ \rho $ and with zero pressure. For intermediate values
of $ R $ the form of $ \rho $ is clearly more complicated but a formula
of the type

\begin{equation}
\rho = \left( u_R \; R + {{d} \over R} + s \right) {1 \over
{R^{D-1}}} \label{rogen}
\end{equation}
where
\begin{eqnarray}
\lim_{R\to\infty} u_R = \cases{ 0 \quad & {\rm FRW } \cr
  u_{\infty} \neq 0 & {\rm Inflationary } \cr}
\end{eqnarray}
 This equation of state is qualitatively
correct for all $ R $ and becomes exact for $ R \to 0 $ and $ R \to
\infty $ . The parameters
$u_R , d$ and $ s $ are positive constants and the $u_R$
varies smoothly with $R$.

The pressure associated to the energy density (\ref{rogen}) takes then
the form
\begin{equation}
p  = {1 \over {D-1}} \left( {d \over R} -  u_R \; R\right) {1 \over
{R^{D-1}}} \label{pgen}
\end{equation}

Inserting eq.(\ref{rogen}) into the Einstein-Friedmann equations
[eq.(\ref{eins})] we find
\begin{equation}
{1 \over 2}~(D-1)(D-2)~ \left({{d R}\over{dT}}\right)^2  =  \left( u_R \; R +
 {d \over R} + s \right) {1 \over
{R^{D-3}}} \label{enfri}
\end{equation}

We see that $R$ is a monotonic function of the cosmic time $T$.
Eq.(\ref{enfri}) yields

\begin{equation}
T = \sqrt{{(D-1)(D-2)}\over 2}~
\int_0^R dR \; {{R^{D/2-1}} ~\over {\sqrt{ u_R \; R^2  + d + s\; R}}}
\label{inte}
\end{equation}

where we set $R(0) = 0$.

It is easy to derive the behavior of $R$ for $T \to 0$ and for $T
\to \infty$.

For  $T \to 0$, $ R  \to 0$, the term  $d/R$ dominates in
eq.(\ref{enfri}) and
\begin{equation}
R(T) ~ \buildrel{T \to 0}\over \simeq ~ {D \over 2}
\left[{{2 d}\over {(D-1)(D-2)}}\right]^{1 \over D}~ T^{2 \over D}
\label{rcer}
\end{equation}
For $T \to \infty$, $R\to \infty$ and  the term $u_R \,R$ dominates  in
eq.(\ref{enfri}). Hence,
\begin{equation}
R(T) ~ \buildrel{T \to \infty}\over \simeq ~
\left[{{(D-2) u_{\infty}}\over {2(D-1)}}\right]^{1 \over {D-2}}~
T^{2 \over {D-2}}
\label{rinf}
\end{equation}
[$u_R$  tends to a constant $u_{\infty}$ for $R\to\infty$].
This expansion is faster than (cold) matter dominated universes where
$ R \simeq  T^{{2 \over {D-1}}} $ . For example, for $ D = 4 $, $
R $ grows linearly with $ T $ whereas for matter dominated universes
$ R \simeq  T^{2/3} $ .
However,  eq.(\ref{rinf}) {\bf is not} a self-consistent solution.
Assuming that
the term $u_R \,R$ dominates for large $R$ we find a scale factor
$ R(T) \simeq  T^{2 \over {D-2}} \simeq \eta^{2 \over {D-4}}$
for $D\neq 4$ and $ R(T) \simeq T \simeq e^{\eta}$ at $ D = 4 $.
This {\bf is not an inflationary universe} but a FRW  universe. The
term $u_R  R$ is absent for large $R$ in FRW universes as explained before.
Therefore, we must  instead use for large $R$
\begin{equation}
\rho = \left( {{d} \over R} + s \right) {1 \over
{R^{D-1}}} \label{rocor}
\end{equation}
Now, for  $T \to \infty , R \to \infty $ and we find a matter dominated
regime:
\begin{equation}
R(T)  \buildrel{T \to \infty}\over \simeq
\left[{{(D-1) s}\over {2(D-2)}}\right]^{1 \over {D-1}}~ T^{2 \over {D-1}}
\label{mdom}
\end{equation}

For intermediate values of $T~ ,~ R(T)$ is a continuous and
monotonically increasing function of $T$.

In summary, the universe starts at $ T = 0 $ with a singularity of
the type dominated by radiation. (The string behaviour for $R \to 0 $
is like usual radiation).
Then, the universe expands monotonically,
growing for large $ T $ as $ R \simeq  T^{{2 \over {D-1}}} $ .
In particular, this gives $ R \simeq  T^{2/3} $ for $ D = 4 $.

It must be noticed that the qualitative
form of the solution $R(T)$ does not depend on the particular
positive values of $ u_R , d $ and $ s $.

We want to stress that we achieve a {\bf self-consistent} solution of the
Einstein equations with string sources since the  behaviour
of the string pressure and density given by eqs.(\ref{rogen})-(\ref{pgen})
precisely holds in universes with power like $ R(T) $.

In ref.\cite{cos} similar results were derived using arguments based
on the splitting of long strings.

\subsection{\bf Thermodynamics of strings in cosmological spacetimes}

Let us consider a comoving volume $ R^{D-1} $ filled by a gas of
strings. The entropy change for this system is given by:
\begin{equation}
{\cal T} dS = d (\rho \; R^{D-1}) ~+ ~ p\;d(R^{D-1})
\label{dentr}
\end{equation}
The continuity equation (\ref{cont}) and (\ref{dentr}) implies that $ dS/dT $
 vanishes.  That is, the entropy per comoving volume stays constant in
time.
 Using now the thermodynamic relation \cite{wei}
\begin{equation}
{{dp}\over{d{\cal T}}} = {{p+\rho}\over {\cal T}}
\end{equation}
it follows \cite{romi} that
\begin{equation}
S = {{R^{D-1}}\over {\cal T}}(p + \rho) + {\rm constant}
\label{entro}
\end{equation}

Eq.(\ref{entro}) together with  eqs.(\ref{rogen}) and (\ref{pgen}) yields
the temperature as a function of the expansion factor $ R $. That is,
\begin{equation}
{\cal T} = {1 \over S}\left\{ s + {1 \over {D-1}} \left[ {{D\; d}\over R} +
(D -2)\; u_R \; R \right] \right\}
\label{tempe}
\end{equation}
where $ S $ stands for the (constant) value of the entropy.

Eq.(\ref{tempe}) shows that for small $ R ,~ {\cal T} $ scales as  $ 1/R $
whereas for large $ R $ it scales as $ R $. The small $ R $ behaviour
of $ {\cal T} $ is the usual exhibited by radiation.

For large $ R $, in FRW universes $u_R \to  0 $ and  the constant term
in $s$ dominates. We just find a cold matter behaviour for large $R$.

For large $ R $ in inflationary universes, $u_R \to  u_{\infty}$
 and eq.(\ref{tempe}) would indicate a temperature that {\bf
grows} proportionally to $R$. However, as stressed in ref.\cite{cos},
the decay of long strings  (through splitting) makes $u_R$
exponentially decreasing  with $R$.

\section{\bf Effective String  Equations with the String Sources Included}

Let us consider now the cosmological equations obtained from the low
energy string effective action including the string matter as a
classical source. In D spacetime dimensions, this action can be
written as
\begin{eqnarray}
S & = & S_1 + S_2 \cr
S_1 & = &  {1 \over 2} \int d^Dx \; \sqrt{-G} ~e^{-\Phi}~\left[\; R +
G_{AB}\; \partial^A\Phi~\partial^B\Phi~+ 2~U(G,\Phi) - c\;\right] \cr
S_2 & = &  -{1 \over {4\pi\alpha'}}\sum_{strings}\int d\sigma d\tau~
G_{AB}(X)~\partial_{\mu}X^A\;\partial^{\mu}X^B \qquad,
\label{accef}
\end{eqnarray}
Here $ A, B = 0, \ldots , D-1 $.
This action is written in the so called `Brans-Dicke frame' (BD)
or `string frame', in which matter couples to the metric tensor in the
standard way. The BD frame metric coincides with the sigma model
metric to which test strings are coupled.

Eq.(\ref{accef}) includes the dilaton field ($\Phi$) with a
potential $U(G,\Phi)$ depending on the dilaton and graviton
backgrounds; $ c $ stands for the central charge deficit or
cosmological constant term. The antisymmetric tensor field was not
included, in fact it is irrelevant   for the results obtained here.
Extremizing the action (\ref{accef}) with respect to $G_{AB}$ and $\Phi$
yields the equations of motion
\begin{eqnarray}
R_{AB} + \nabla_{AB}\Phi  + 2 \, {{\partial U}\over {\partial
G_{AB}}} - {{G_{AB}}\over 2} \left[ R + 2 \, \nabla^2\Phi -
(\nabla \Phi)^2 - c + 2 \, U \right] & = & e^{\phi}~T_{AB} \nonumber \\ \cr
R + 2 \, \nabla^2\Phi - (\nabla \Phi)^2 - c + 2 U -
{{\partial U}\over {\partial \Phi}} & = & 0 ~,
\label{eqef}
\end{eqnarray}
which can be more simply combined as
\begin{eqnarray}
R_{AB} + \nabla_{AB}\Phi  + 2 \; {{\partial U}\over {\partial
G_{AB}}} - G_{AB}\;
{\,{\partial U}\over {\partial \Phi}}  & = & e^{\Phi}~T_{AB}
\nonumber \\ \cr
R + 2 \; \nabla^2\Phi - (\nabla \Phi)^2 - c + 2 \, U -
{{\partial U}\over {\partial \Phi}} & = & 0
\label{eqeff}
\end{eqnarray}
Here $ T_{AB} $ stands for the energy momentum tensor of the strings
as defined by eq.(\ref{tens}).
It is also convenient to write these equations as
\begin{equation}
R_{AB}- {{G_{AB}}\over 2}\;R =   T_{AB} + \tau_{AB}
\end{equation}
where $ \tau_{AB} $ is the dilaton  energy momentum tensor :
\begin{eqnarray}
\tau_{AB} & = &  -\nabla_{AB}\Phi + {{G_{AB}}\over 2}
 \left[ 2 \, {{\partial U}\over {\partial \Phi}} - R \right] \nonumber
\end{eqnarray}
The Bianchi identity
\begin{eqnarray}
\nabla^A\left(R_{AB}- {{G_{AB}}\over 2}\;R \right)  & = & 0
\nonumber \cr
\end{eqnarray}
yields, as it must be, the conservation equation,
\begin{equation}
\nabla^A\left( T_{AB} + \tau_{AB}\right) = 0
\end{equation}
It must be noticed that eqs.(\ref{eqeff}) do not reduce to the
Einstein equations of General Relativity even when $ \Phi = U = 0 $.
Eqs. (\ref{eqeff}) yields in that case the
Einstein equations {\it plus} the condition $ R = 0 $.

\subsection{Effective String Equations in Cosmological Universes}

For the homogeneous isotropic spacetime
geometries described by eq.(\ref{met}) we  have
\begin{eqnarray}
R_0^0 & = & - (D-1) ( {\dot H} + H^2 ) \cr
R_i^k & = & - \delta_i^k ~ [ {\dot H} + (D-1)\,H^2 ] \cr
R  & = & - (D-1) ( 2\;{\dot H} + D\,H^2 ) .
\end{eqnarray}
where $H \equiv {1 \over R}\,{{dR}\over{dT}}$ .

The equations of motion (\ref{eqeff}) read
\begin{eqnarray}
{\ddot \Phi } - (D-1) ( {\dot H} + H^2 ) - {{\partial U}\over
{\partial \Phi}} & = &  e^{\Phi}~\rho  \nonumber \\ \cr
 {\dot H} + (D-1)\,H^2 - H \, {\dot \Phi }  + {{\partial U}\over
{\partial \Phi}} + { R \over {D-1}}{{\partial U}\over
{\partial R}} & = &  e^{\Phi}~p  \nonumber \\ \cr
2\;{\ddot \Phi } + 2(D-1)\,H \, {\dot \Phi }- {\dot \Phi }^2
 - (D-1) ( 2\;{\dot H} + D\,H^2 ) - 2\; {{\partial U}\over
{\partial \Phi}} - c + 2\, U  & = & 0
\label{eqcos}
\end{eqnarray}
where dot $ {}^. $ stands for $ {{d}\over{dT}}$, and
\begin{equation}
\rho = T_0^0  \qquad , \qquad -\delta_i^k~p = T_i^k ~.
\end{equation}
The conservation equation takes the form of eq.(\ref{cont})
\begin{equation}
\dot{\rho} + (D-1)\, H\, (p + \rho ) = 0 ~~.
\end{equation}
By defining,
\begin{eqnarray}
\Psi & \equiv & \Phi - \log{\sqrt{-G}} = \Phi - (D-1)\,\log R \cr
{\bar \rho } & =  & e^{\Phi}~\rho \quad, \quad {\bar p } =  e^{\Phi}~p ~,
\label{psiba}
\end{eqnarray}
 equations (\ref{eqcos}) can be expressed in a more compact form as
\begin{eqnarray}
{\ddot \Psi } - (D-1)\,  H^2  - \left.{{\partial U}\over
{\partial \Psi}}\right|_R & = &  {\bar \rho }  \nonumber \\ \cr
 {\dot H}  - H \, {\dot \Psi }  +  \left.{ R \over {D-1}}{{\partial U}\over
{\partial R}}\right|_{\Psi} & = &    {\bar p }  \nonumber \\ \cr
{\dot \Psi }^2 - (D-1)\,H^2  - 2\; {\bar \rho } - 2\, U  + c & = & 0
\; ,
\label{eqcof}
\end{eqnarray}
The conservation equation reads
\begin{equation}
\dot{{\bar\rho}} -  {\dot \Psi }\; {\bar \rho }+ (D-1)\, H\,  {\bar p }
= 0
\end{equation}

As is known, under the duality transformation $ R \longrightarrow
R^{-1} $ , the dilaton transforms as $\Phi \longrightarrow
\Phi + (D-1)\,\log R $. The shifted dilaton $\Psi$ defined by
eq.(\ref{psiba}) is invariant under duality.

The transformation
\begin{equation}
 R' \equiv R^{-1} \quad, ~
\end{equation}
implies
\begin{equation}
\Psi' = \Psi \quad, ~ H' = -H \quad, ~ {\bar p' }=-p \quad, ~
{\bar\rho'} = {\bar\rho}
\end{equation}
provided $ u = d $, that is, a   duality invariant  string source.
This is the duality  invariance transformation of eqs.(\ref{eqcof}).

Solutions to the effective string equations have been extensively
treated in the literature \cite{eqef} and they are not our main
purpose. For the sake of completeness, we briefly analyze the limiting
behaviour of these equations for $ R \to \infty$ and $ R \to 0 $.

It is difficult to make a complete analysis of the  effective string
equations (\ref{eqcof}) since the knowledge about the potential $ U $
is rather incomplete. For weak coupling ($ e^{\Phi} $ small ) the
supersymmetry breaking produces an effective potential that decreases
very fast (as the exponential of an exponential of $\Phi$) for
 $\Phi \to -\infty$.

Let us analyze the asymptotic behavior of  eqs.(\ref{eqcof}) for $ R
\to \infty $ and  $ R \to 0 $ assuming that the potential $ U $
can be ignored.  It is easy to see that a power behaviour Ansatz both
for $ R $ and for $ e^{\Psi} $ as functions of $ T $ is consistent
with these equations. It turns out that the string sources do not
contribute to the leading behaviour here, and we find for  $ R \to 0 $
\begin{eqnarray}
R_{\mp} =&  C_1 \; T^{\pm 1/\sqrt{D-1}} ~ \to 0\quad,\cr
 e^{\Psi_{\mp}} =& C_2 \; T^{-1} ~ \to  \left\{
\begin{array}{ll}  \infty \\  0 \end{array} \right.
\label{solef}
\end{eqnarray}
Where $ C_1 $ and $  C_2 $ are constants.
Here the branches $(-)$ and $(+)$ correspond to $ T \to 0 $ and to
 $ T \to \infty $ respectively. In both regimes $ R_{\mp} \to 0 $
and $ e^{\Phi_{\mp}} \to 0$.

The potential $ U(\Phi) $ is hence negligible in these regimes. In
terms of the conformal time $ \eta $ , the behaviours (\ref{solef})
result
\begin{eqnarray}
R_{\mp} =&  C_1' \; \eta^{\pm {1 \over {\sqrt{D-1} \mp 1}}} \to 0 \cr
 e^{\Psi_{\mp}} =& C_2' \; \eta^{-{{\sqrt{D-1}}\over{\sqrt{D-1} \mp
1}}} \to  \left\{
\begin{array}{ll}  \infty \\  0 \end{array} \right.
\label{sefd}
\end{eqnarray}
Where $ C_1' $ and $  C_2' $ are constants.
The branch $(-)$ would describe an expanding non-inflationary
behaviour near the initial singularity $ T = 0 $ , while the branch
$(+)$ describes a `big crunch' situation and is rather unphysical.

 Similarly,   for   $ R \to \infty $ and
 $ e^{\Phi} \to \infty $, we find
\begin{eqnarray}
R_{\mp} =&  D_1 \; T^{\mp 1/\sqrt{D-1}} ~ \to \infty \quad,\cr
 e^{\Psi_{\mp}} =& D_2 \; T^{-1} ~ \to  \left\{
\begin{array}{ll}  \infty \\  0 \end{array} \right.
\label{salef}
\end{eqnarray}
Where $ D_1 $ and $  D_2 $ are constants.
Here again, the branches $(-)$ and $(+)$ correspond to $ T \to 0 $ and to
 $ T \to \infty $ respectively, but now in both regimes
$ R_{\mp} \to \infty$ and $ e^{\Phi_{\mp}} \to \infty $. (In this
limit, one is not guaranteed that $ U $ can be consistently
neglected). In terms of the conformal time, eqs.(\ref{salef}) read
\begin{eqnarray}
R_{\mp} =&  D_1' \; \eta^{\mp {1 \over {\sqrt{D-1} \pm 1}}} \to \infty \cr
 e^{\Psi_{\mp}} =& D_2' \; \eta^{-{{\sqrt{D-1}}\over{\sqrt{D-1} \pm
1}}} \to  \left\{
\begin{array}{ll}  \infty \\  0 \end{array} \right.
\label{safd}
\end{eqnarray}
The branch $(+)$ describes a noninflationary expanding behaviour for $
T \to \infty $ faster than the standard matter dominated expansion,
while the branch $(-)$ describes a super-inflationary behaviour
$\eta^{-\alpha}$, since $ 0 < \alpha < 1 $, for all D.

The behaviours (\ref{solef}) for  $ R_{\mp} \to 0 $ and (\ref{salef})
for $ R_{\mp} \to \infty$ are related by duality $ R \leftrightarrow 1/R $.

\subsection{\bf String driven inflation?}

Let us consider now the question of whether de Sitter spacetime may be
a self-consistent solution of the effective string equations
(\ref{eqcos})
with the string sources included. The strings in cosmological
universes like de Sitter spacetime have the equation of state
(\ref{rogen})-(\ref{pgen}).
Since $ e^{\Psi} = e^{\Phi} \; R^{1-D} $ :
\begin{eqnarray}
{\bar \rho} & = & e^{\Psi} \left( u \; R + {d \over R} + s \right)
\label{denb} \\ \cr
{\bar p} & = &
{{e^{\Psi}} \over {D-1}} \left( {d \over R} -  u \; R\right)
 \label{prba}
\end{eqnarray}

In the absence of dilaton potential and cosmological constant term,
the string sources do not generate de Sitter spacetime as discussed in
sec. V.A. We see that for $ U = c = 0 $ , and $ R = e^{H T} $ ,
eqs.(\ref{eqcof}) yields to a contradiction
 (unless $ D = 0 $ ) for
the value of $\Psi $ , required to be
 $ -H  T \, + $ constant.

A self-consistent solution describing asymptotically de Sitter
spacetime self-sustained by the string equation of state
(\ref{denb})-(\ref{prba}) is given by
 \begin{eqnarray}
 R & = & e^{H T} ~~, ~~ H = {\rm  constant} > 0 ~~,\cr
 2U-c & = & D\; H^2
=  {\rm  constant} \cr
\Psi_{\pm} & = & \mp H T
\pm i\pi + \log{{(D-1)\,H^2}\over {\rho_{\pm}}} \cr
\rho_+ & \equiv & u \quad , \quad \rho_- \equiv d \quad
\label{anti}
\end{eqnarray}
The branch $\Psi_+$ describes the solution for $ R \to \infty $ ( $ T
\to + \infty  ) $, while the branch  $\Psi_-$ corresponds to  $ R \to 0 $
 ( $ T \to - \infty  ) $. De Sitter spacetime with lorentzian
signature self-sustained by the strings necessarily requires a constant
imaginary piece $ \pm i \pi $ in the dilaton field. This makes $
e^\Psi < 0 $ telling us that the gravitational constant $ G \sim
 e^\Psi < 0 $ here describes antigravity.

Is interesting to notice that in the euclidean signature case, i. e.
(+++\ldots++), the Ansatz ${\dot H} = 0,~ 2U-c=$constant, yields a
constant curvature geometry with a real dilaton, but which is of
Anti-de Sitter type. This solution is obtained from
eqs.(\ref{prba})-(\ref{anti}) through the transformation
\begin{equation}
{\hat X}^0 = i T ~~~,~~~{\hat H} = -i H  ~~~,~~~ X^i = X^i ~~~
,~~~\Psi = \Psi
\label{traf}
\end{equation}
which maps the Lorentzian de Sitter metric into the positive definite
one
\begin{equation}
d{\hat s}^2 = (d{\hat X}^0)^2 + e^{{\hat H}{\hat X}^0}~(d\vec{X})^2.
\label{posi}
\end{equation}
The equations of motion (\ref{eqcof}) within the constant curvature
Ansatz $({\dot{\hat H}} = {\ddot \Psi } = 0 )$
are mapped onto the equations
\begin{eqnarray}
 (D-1)\, {\hat H}^2  - \left.{{\partial U}\over
{\partial \Psi}}\right|_R & = &  {\bar \rho }  \nonumber \\ \cr
  {\hat H} \, {{d \Psi}\over {d {\hat X}^0 }}
+  \left.{ R \over {D-1}}{{\partial U}\over
{\partial R}}\right|_{\Psi} & = &    {\bar p }  \nonumber \\ \cr
-({{d \Psi}\over {d {\hat X}^0 }})^2
+(D-1)\,{\hat H}^2  - 2\; {\bar \rho } - 2\, U  + c & = & 0
\; ,
\label{eqhat}
\end{eqnarray}
with the solution
\begin{eqnarray}
 R & = & e^{{\hat H} {\hat X}^0} ~~, ~~ {\hat H} = {\rm  constant} > 0 ~~,\cr
 c -2 \, U & = & D\; {\hat H}^2
=  {\rm  constant} \cr
\Psi_{\pm} & = & \mp {\hat H } {\hat X}^0
+ \log{{(D-1)\,
{\hat H}^2}\over {\rho_{\pm}}} \cr
\rho_+ & \equiv & u \quad , \quad \rho_- \equiv d \quad
\label{antit}
\end{eqnarray}
Both solutions (\ref{antit}) and (\ref{anti}) are mapped one into
another through the transformation (\ref{traf}).

\bigskip

It could be recalled that in the context of (point particle) field
theory, de Sitter spacetime (as well as anti-de Sitter) emerges as an
exact selfconsistent solution of the semiclassical Einstein equations
with the back reaction included \cite{uno} - \cite{dos}. (Semiclassical in this
context, means that matter fields including the graviton are quantized
to the one-loop level and coupled to the (c-number) gravity background
through the expectation value of the energy-momentum tensor $T_A^B$ . This
expectation value is given by the trace anomaly:
$ <T_A^A> = {\bar \gamma} \; R^2  $). On the other hand,
the $\alpha'$ expansion of the effective string action admits anti-de
Sitter spacetime (but not  de Sitter) as a solution when the quadratic
curvature corrections (in terms of the Gauss-Bonnet term) to the
 Einstein action are included \cite{tres}. It appears that the
corrections to the anti-de Sitter constant curvature are qualitatively
similar in the both cases, with $\alpha'$ playing the r\^ole of the trace
anomaly parameter ${\bar \gamma}$\cite{dos}.

The fact that de Sitter inflation with true gravity $ G \sim
 e^\Psi > 0 $ does not emerge as a solution of the effective string
equations does not mean that string theory excludes inflation. What
means is that the effective string equations are not  enough to get inflation.
The effective string action is a low
energy field theory approximation to string theory containing only the
{\it massless} string modes ({\it massless} background fields).

The vacuum energy scales to start inflation (physical or true vacuum)
are typically of the order of the Planck mass \cite{romi} - \cite{linde}
where the effective string action approximation breaks
down. One must consider the massive string modes (which are absent
from the effective string action) in order to properly get the
cosmological condensate yielding de Sitter inflation. We do not have
at present the solution of such problem.

\bigskip

\bigskip

\begin{centerline}
{TABLE 4. Asymptotic solution of the string effective equations}
(including the dilaton).

\bigskip

\bigskip

{\bf EFFECTIVE STRING EQUATIONS}

{\bf  SOLUTIONS IN COSMOLOGY}

\vskip 20pt
\begin{tabular}{|l|l|l|}\hline
Effective String &  $~~R(T)\to 0$  &  $~~R(T)\to \infty$\\
\hspace{4mm}equations & \hspace{4mm}behaviour &
\hspace{4mm} behaviour \\ \hline
$ $&  $ $ & $ $ \\
\hspace{4mm}$ T \to 0 $ & $~\simeq\, T^{+ 1/\sqrt{D-1}}$ &
$~\simeq\,T^{-1/\sqrt{D-1}}$\\
$ $&  $ $ & $ $ \\ \hline
$ $&  $ $ & $ $ \\
\hspace{4mm}$ T\to \infty$ & $~\simeq \, T^{-1/\sqrt{D-1}}$ &
$~\simeq \, T^{+ 1/\sqrt{D-1}}$ \\
$ $&  $ $ & $ $ \\ \hline
\end{tabular}
\end{centerline}

\section{Multi-Strings and Soliton Methods in de Sitter Universe}

Among the cosmological backgrounds, de Sitter
spacetime occupies a special
place. This is, in one hand  relevant for inflation and on the
other hand string propagation turns to be specially
interesting  there \cite{dvs87} - \cite{igor}.
String unstability, in the sense that the string proper
length grows indefinitely is particularly present in de Sitter.
The string dynamics in de Sitter universe is described by a
generalized sinh-Gordon model with a potential  unbounded from
below \cite{prd}. The sinh-Gordon function $\alpha(\sigma,\tau)$
having a clear physical meaning : $ H^{-1} e^{\alpha(\sigma,\tau)/2} $
determines the string proper length.
Moreover the classical string equations of motion (plus the
string constraints) turn to be integrable in de Sitter universe
\cite{prd,dms}. More precisely, they are equivalent to a
non-linear sigma model on the grassmannian $SO(D,1)/O(D)$ with periodic
boundary conditions (for closed strings). This sigma model
has an associated linear system \cite{zm1} and using it, one can show the
presence  of an infinite number of conserved quantities \cite{dv78}.
In addition, the string constraints imply
a zero energy-momentum tensor and these constraints
are compatible with the integrability.

The so-called dressing method \cite{zm1} in soliton theory
allows to construct solutions of non-linear classically
integrable models using the associated linear
system. In ref.\cite{cdms}  we systematically construct string solutions
in three dimensional de Sitter spacetime.
We start from a given exactly known solution
of the string equations of motion and constraints in de Sitter \cite{dms}
and then we ``dress'' it.
The string solutions reported there indeed apply to cosmic strings
in de Sitter spacetime as well.

 The invariant interval in $D$-dimensional de Sitter space-time is given by
\begin{equation}
ds^2=dT^{2} ~-~ \exp[2HT]~ \sum_{i=1}^{D-1} (dX^{i})^2 .
\end{equation}
 Here $T$ is the so called cosmic time. In terms of the conformal time
$\eta$,
$$
\eta \equiv-{{\exp[-H T]} \o H} \quad , -\infty < \eta \leq 0 ~ ,
$$
 the line element becomes
$$
ds^2 = {1 \o {H^{2} \eta^{2}}}[  d\eta^2 ~-~ \sum_{i=1}^{D-1}
(dX^{i})^{2}\, ]  ~.
$$
 The de Sitter spacetime can be considered as a $D$-dimensional hyperboloid
embedded in a D+1 dimensional flat Minkowski spacetime with coordinates
$(q^0,...,q^D)$ :
\begin{equation}\label{mehip}
ds^2 = {1\o {H^2} }[ - (dq^0)^2 ~+~ \sum_{i=1}^{D} (dq^i)^2 \, ]
\end{equation}
 where
\begin{eqnarray}
q^0 &=& ~\sinh{H T}~ + {{H^2} \o 2} \exp[H T]~ \sum_{i=1}^{D-1}
(X^i)^2 ~~, \cr \cr
q^1 &=& ~\cosh{H T}~ - {{H^2} \o 2} \exp[H T]~ \sum_{i=1}^{D-1}
(X^i)^2 ~~ , \cr \cr
q^{i+1} &=& ~H \exp[H T]~ X^{i} \quad , \quad  1 \leq i \leq D-1, ~
-\infty < T , ~ X^{i} < + \infty .
\end{eqnarray}
 The complete de Sitter manifold is the hyperboloid
$$
-(q^0)^2 + \sum_{i=1}^{D} (q^i)^{2} = 1.
$$
 The coordinates  $(T, X^{i})$  and  $(\eta , X^{i})$  cover only
the half of the de Sitter manifold $q^0 + q^1 > 0$.

 We will consider a string propagating in this D-dimensional
space-time.The string equations of motion (\ref{conouno}) in the metric
(\ref{mehip}) take the form:
\begin{equation}
\partial_{+-}q + (\partial_{+}q . \partial_{-}q)\;q = 0 \quad
{\rm with} \quad q.q = 1 ,
\label{desmo}
\end{equation}
where . stands for the Lorentzian scalar product $a.b \equiv -a_0b_0 +
\sum_{i=1}^{D}a_ib_i ~ ,~x_\pm \equiv {1\o{2}}(\tau \pm \s)$ and
$\partial_{\pm}q = {{\partial q} \o {\partial x_{\pm}}}$.
The string constraints (\ref{conodos}) become for de Sitter universe
\begin{equation}
T_{\pm\pm}= {\partial q \o {\partial x_\pm}}.
{\partial q \o {\partial x_\pm}} = 0           . \label{bincu}
\end{equation}

 Eqs.(\ref{desmo}) describe a non compact O(D,1) non-linear sigma model in
two dimensions. In addition, the (two dimensional) energy-momentum
tensor is required to vanish by the constraints eqs.(\ref{bincu}) .
This system of non-linear partial differential equations can be
reduced by choosing an appropriate basis for the string coordinates
in the ($D+1$)-dimensional Minkowski space time $(q^0,...,q^D)$
to a noncompact Toda model \cite{prd}.

These equations can be rewritten in the form of a chiral field model
on the Grassmanian
$G_{D}=SO(D,1)/O(D)$. Indeed, any element  ${\bf g}\in G_{D}$ can be
parametrized with a real vector $q\rangle $ of the unit pseudolength
\begin{equation}
{\bf g}=1-2 |q\rangle\langle q|J,\ \\\\\ \  \langle q|J|q\rangle =1.
\label{G}
\end{equation}
In terms of $\bf g$, the string equations (\ref{desmo}) have the following form
\begin{equation}
2\,{\bf g}_{\xi \eta }-{\bf g}_{\xi }\,{\bf g\, g}_{\eta }-
{\bf g}_{\eta }\,{\bf g\, g}_{\xi }=0~,
\label{geq}
\end{equation}
and  the conformal constraints  (\ref{bincu}) become
\begin{equation}
\tr \, {\bf g}_{\xi }^{2}=0,\ \ \ \tr \, {\bf g}_{\eta }^{2}=0~.
\label{gcond}\end{equation}

The fact that ${\bf g}\in G_{D}$ implies that $\bf g $ is a
real matrix with the following properties:

\begin{equation}
{\bf g}=J{\bf g} ^{\mbox{t}}J,\ \ \ {\bf g} ^{2}=I,\ \ \ \
tr \, {\bf g} =D-1 \ , \ \ {\bf g} \in SL(D+1, R).
\label{gcond1}\end{equation}
These conditions are equivalent to the existence of the  representation
(\ref{G}).
Equation (\ref{geq}) is the compatibility condition for
the following overdetermined linear system:
\begin{equation}
\Psi _{\xi}=\frac{U}{1-\lambda }\Psi ,\ \
\Psi _{\eta}=\frac{V}{1+\lambda }\Psi ,
\label{psieq}
\end{equation}
where
\begin{equation}
U={\bf g}_{\xi}\, {\bf g},\ \ \ V={\bf g}_{\eta}\, {\bf g} \ \ .
\label{UV}\end{equation}
Or, in terms of the vector $q\rangle $
$$
U=2\,  q_{\xi }\rangle\langle q \,
J -2 \, q\rangle \langle q_{\xi }\, J , \ \
$$
$$
V=2\,  q_{\eta }\rangle\langle q \, J-2 \,
q\rangle \langle q_{\eta }\, J .
$$
Eq. (\ref{G}) can be easily inverted yielding $q$ in terms of the
matrix $g$:
\begin{equation}
q_0 = \sqrt{{g_{00} -1}\over{2}} \quad , \quad
q_i = \sqrt{{1 - g_{ii} }\over{2}} \; \; 1 \leq i \leq D \;({\rm
no~sum ~ over~}i)
\label{qG}
\end{equation}

The use of overdetermined linear systems to solve non-linear
partial differential equations associated to them goes back to refs.
\cite{ist}. (See refs.\cite{sol} -\cite{lib2} for further references).

In order to fix the freedom in the definition
of $\Psi $ we shall identify
\begin{equation}
\Psi (\lambda =0)=\bf g.
\label{l0g}\end{equation}

This condition  is compatible with the above equations since the matrix
function $\Psi $ at the point $\lambda =0$ satisfies
the same equations as $\bf g$. Thus the problem of
constructing exact solutions of the string equations
is reduced to finding  compatible solutions of the linear equations
(\ref{psieq}) such that  ${\bf g}=\Psi(\lambda =0)$ satisfies
the constraints eqs.(\ref{gcond}) and (\ref{gcond1}).

We concentrate below on the linear system (\ref{psieq}) since this is the
main tool to derive new string solutions in de Sitter spacetime.

In ref.\cite{cdms} the dressing method was applied as follows. We started
from the exact
ring-shaped string solution $ q_{(0)} $ \cite{dms} and we find the
explicit solution $ \Psi^{(0)}(\lambda) $ of the
associated linear system, where $\lambda$ stands for the spectral
parameter.
Then, we propose a new solution $ \Psi(\lambda) $ that differs from
$ \Psi^{(0)}(\lambda) $ by a matrix rational in $\lambda$.
Notice that  $ \Psi(\lambda = 0) $ provides in general a new string solution.

We then show that this rational matrix must have at least {\bf four} poles,
 $ \lambda _{0}, 1/\lambda _{0}, \lambda _{0}^* , 1/\lambda _{0} ^* $,
as a consequence of the symmetries of the problem.
The residues of these poles are shown to be one-dimensional
projectors. We then prove that these projectors are formed by vectors
which can all be expressed in terms of an arbitrary complex constant
vector $ |x_0 \rangle $ and the complex parameter $ \lambda _{0} $.
This result holds for arbitrary starting solutions $ q_{(0)} $.

Since we consider closed strings, we impose a $2\pi$-periodicity on
the string variable $ \sigma $ . This restricts $ \lambda _{0} $ to
take discrete values that we succeed to express in terms
of  Pythagorean numbers.
In summary, our solutions depend on two arbitrary complex numbers
contained in  $ |x_0 \rangle $ and two integers $ n $ and $ m $ . The
counting of degrees of freedom is analogous to $2 + 1$ Minkowski
spacetime except that left and right modes are here mixed up  in a
non-linear and precise way.

The vector $ |x_0 \rangle $ somehow indicates the polarization of the
string. The integers $(n,m)$  determine the string winding. They fix
the way in which the string winds around the origin in the spatial
dimensions (here $S^2$ ). Our starting solution $q_{(0)}(\sigma,
\tau)$ is a stable string
winded $n^2 + m^2$ times  around the origin in de Sitter space.

The matrix multiplications involved in the computation of
the final solution were done with the help of  the computer
program of symbolic calculation ``Mathematica''. The resulting solution
$ q(\sigma, \tau) = (q^0, q^1, q^2, q^3) $ is a complicated
combination of trigonometric functions of $ \sigma $ and hyperbolic
functions of $\tau$. That is, these string solitonic solutions
do not oscillate in time. This is a typical feature of string
unstability \cite{dms} - \cite{gsv} - \cite{agn}. The new feature here is
that strings (even stable solutions) do not oscillate neither
for $ \tau \to 0 $, nor for $\tau \to \pm\infty $.

We plot in figs. 1-7 the solutions for significative values
of  $|x_0 \rangle $ and $(m, n)$ in terms of the comoving coordinates
$( T, X^1, X^2 )$
\begin{equation}
T = \frac{1}{H}\log(q^0 + q^1) ~~,~~~ X^1 =\frac{1}{H}
\frac{q^2}{q^0 + q^1} ~~,~~X^2 =\frac{1}{H}
\frac{q^3}{q^0 + q^1} ~~
\label{cooXT}
\end{equation}
The first feature to point out is that our solitonic solutions
describe {\bf multiple} (here five  or three) strings,
as it can be seen from the fact that for a given time $T$
 we find several different values for $\tau$.
That is, $\tau$ is a {\bf multivalued} function of $T$
for any fixed $\sigma$ (fig.1-2). Each branch of $\tau$
as a function of $T$  corresponds to a different string.
This is a entirely new feature for strings in curved spacetime, with
no analogue in flat spacetime where the time coordinate
can always be chosen proportional to $\tau$.
In flat spacetime, multiple string solutions are described
by multiple world-sheets.
Here, we have a {\bf single} world-sheet describing
several independent and simultaneous strings as a consequence of the
coupling with the spacetime geometry.
Notice that we consider {\it free} strings. (Interactions
among the strings as splitting or merging  are not considered).
Five  is the generic number of strings in our dressed
solutions. The value five can be related to the fact
that we are dressing a one-string solution ($ q_{(0)} $) with {\it four}
poles. Each pole adds here an unstable string.

In order to describe the real physical evolution, we eliminated
numerically $ \tau = \tau (\sigma, T) $ from the solution and expressed
the spatial comoving coordinates $X^1$ and $X^2$ in terms of
$T$ and $\sigma$.

We plot $\tau(\sigma, T)$ as a function of $\sigma$ for different fixed values
of $T$ in fig.3-4. It is a sinusoidal-type function. Besides the
customary closed string period $2\pi$, another period appears
which varies on $\tau$. For small $\tau $ , $\tau = \tau(\sigma, T)$ has
a convoluted shape while for larger $ \tau $ (here $  \tau \leq 5
$), it becomes a regular sinusoid. These behaviours reflect very
clearly in the evolution of the spatial coordinates and shape
of the string.

The evolution of the five (and three) strings
simultaneously described by our solution as a function
of $T$, for positive $T$
 is shown in figs. 5-7. One string is stable (the 5th one). The other four are
unstable. For the stable string, $(X^1, X^2)$ contracts in time
precisely as $ e^{-HT} $, thus keeping the proper amplitude
$(e^{HT} X^1, e^{HT} X^2)$ and proper size constant.
For this stable string $(X^1, X^2) \leq {1 \over H}$.
($1/H$ = the horizon radius).  For the other
(unstable) strings, $(X^1, X^2)$ become very fast constant in time,
the proper size expanding as the universe itself like  $ e^{HT} $ .
For these strings $(X^1, X^2) \geq {1 \over H}$.
These exact solutions display remarkably the asymptotic string
behaviour found in refs.\cite{prd,gsv}.

In terms of the sinh-Gordon description, this means that for the
strings outside the horizon the sinh-Gordon function
$\alpha(\sigma,\tau)$
is the same as the cosmic time $T$ up to
a function of $\sigma$. More precisely,
\begin{equation}
\alpha(\sigma,\tau) \buildrel{T >> {1\over H} }\over =
2 H\, T(\sigma, \tau) + \log\left\{2 H^2 \left[ (A^{1}(\sigma)')^2 +
  (A^{2}(\sigma)')^2 \right] \right\} + O(e^{-2HT}).
\label{Tcosa}
\end{equation}
Here $A^1(\sigma)$ and  $A^2(\sigma)$ are the $X^1$ and $X^2$
coordinates outside the horizon. For $T \to +\infty$ these
strings are at the absolute {\it minimum} $\alpha = + \infty $
of the sinh-Gordon potential with infinite size.
The string inside the horizon (stable string) corresponds to the
 {\it maximum} of the potential, $\alpha = 0$.
$ \alpha = 0$ is the only value in which the string
can stay without being
pushed down by the potential to $\alpha = \pm \infty$ and
this also explains why only one stable
string appears (is not possible to put more than one string at
the maximum of the potential without falling down).
These features are {\it generically} exhibited by our
one-soliton multistring solutions, independently of the
particular initial state of the string
(fixed by $|x^0> $ and $(n,m)$).
For particular values of $|x^0> $, the solution describes
three strings, with symmetric shapes from $ T = 0 $, for instance
like a rosette or a circle with festoons (fig. 5-7).

The string solutions presented here trivially embedd on
$D$-dimensional de Sitter spacetime ($D \geq 3$). It must be noticed
that they exhibit the essential physics of strings in D-dimensional
de Sitter universe. Moreover, the construction method used here
works in any number of dimensions.

New classes of multistring solutions in curved spacetime has been recently
found
in \cite{bhln}.

\section{ Strings next to and inside  black holes}

The classical string equations of motion and constraints were solved near the
horizon and near the  singularity of a Schwarzschild black hole in
ref. \cite{negro}. Similar results have been obtained recently in
ref. \cite{nulbh} using the null string approach \cite{nic}.

In  a conformal gauge such that $\tau = 0$ ($\tau$ = worldsheet time
coordinate) corresponds to the
horizon ($r=1$) or to the  black hole  singularity ($r=0$), the
string coordinates express in power series in  $\tau$ near the horizon
and  in power series in $\tau^{1/5}$ around $r=0$.

In ref.\cite{negro}  the string invariant size and the string
energy-momentum tensor were computed. Near the horizon both are finite
and analytic.
Near the  black hole  singularity, the string size, the string energy
and the transverse pressures (in the angular directions) tend to
infinity as $r^{-1}$.
To leading order near  $r=0$, the string behaves as two dimensional
radiation. This  two spatial dimensions are describing the $S^2$ sphere in the
Schwarzschild manifold.

\subsection{String Equations of motion in a Schwarzschild Black Hole.}

The Schwarzschild metric in Schwarzschild coordinates
$(t,r,\theta,\phi)$ takes the following form:
\begin{equation}\label{schsch}
\displaystyle{
ds^2=\left(1-{1\o{r}}\right){\rm d}t^2 - {{{\rm
d}r^2}\o{1-{1\o r}}} - r^2({\rm
d}\theta^2+\sin^2\theta\>{\rm d}\phi^2)\,, }
\end{equation}
where we choose units where  the Schwarzschild radius
$R_s=2m = 1$.

Since we are interested in the whole Schwarzschild manifold and not
just in the external part $r > 1$ where the static Schwarzschild
coordinates are
appropriate, we consider the Kruskal-Szekeres coordinates
$(u,v,\theta,\phi)$ defined by
\begin{equation}\label{dfkru}
u= t_K - r_K \equiv \sqrt{1-r}\; e^{(r-t)/2} \quad , \quad
v= t_K + r_K \equiv \sqrt{1-r}\; e^{(r+t)/2} \; .
\end{equation}
for $v \geq  0, u \geq 0$ and by
\begin{equation}\label{dfkru2}
u= t_K - r_K \equiv -\sqrt{r-1}\; e^{(r-t)/2} \quad , \quad
v= t_K + r_K \equiv \sqrt{r-1}\; e^{(r+t)/2} \; .
\end{equation}
for $v  \geq 0, u \leq 0$. For $v \leq 0$ one just flips the sign of
$v$ in eq.(\ref{dfkru}) or (\ref{dfkru2}) \cite{bibneg}.

The coordinate $ t_K $ is a time-like coordinate, and $ r_K $ spacelike.
In  Kruskal-Szekeres coordinates the Schwarzschild metric takes the form,
\begin{equation}\label{schkru}
ds^2=-{4\over{r}}~e^{-r}\;du\,dv
+r^2~({\rm d}\theta^2+\sin^2\theta\>{\rm d}\phi^2)\,.
\end{equation}
$r$ is a function of the product $uv$ defined by the inverse function
of
$$
uv = [1-r] \, e^r \; .
$$
for $u v \geq 0$. The metric is such coordinates is  regular
everywhere except at its singularity,  $r=0$.

The string equations of motion  in Schwarzschild
coordinates and in the conformal gauge, are
\begin{eqnarray}\label{eqsch}
\rs\ts - \rt\ta + r(r-1)(\tss-\taa)&=&0\,,\cr
{{2r}\over{1-r}}(\rtt-\rss) - {1\over{r^2}}(\ta^2-\ts^2) + 2r ( \tea^2
-\tes^2)&+&\quad\cr
\displaystyle{
2\,r\sin^2\theta\>(\pa^2-\ps^2)} \displaystyle{\,+\,
{1\over{(r-1)^2}}(\rt^2-\rs^2)}&=&0\,.
\end{eqnarray}
\begin{eqnarray}\label{eqang}
r\sin\theta\>(\paa-\pss) + 2\, r\cos\theta\>(\pa\tea-\ps\tes) + 2 \,\sin\theta
\, (\rt\pa-&\rs\ps)=0\,,\cr \cr
r\,(\teaa-\tess)+ 2(\rt\tea-\rs\tes) - r\sin\theta\>\cos\theta\>
(\pa^2-\ps^2)&=0\, .
\end{eqnarray}
The constraints  in Schwarzschild coordinates are
\begin{eqnarray}\label{vinsch}
{{1-r}\over{r}}(\ts^2+\ta^2) + {r\over{r-1}}(\rt^2+\rs^2)
&+& r^2 \left[\tea^2+\tes^2 + \sin^2\theta\>(\pa^2+\ps^2)\right]=0\,,\cr
{{1-r}\over{r}}\ta\ts +  {r\over{r-1}} \rt\rs &+ & r^2\left(\tea\tes
+ \sin^2\theta\>\pa\ps\right)=0\,.
\end{eqnarray}

The string equations of motion in  Kruskal-Szekeres coordinates take
the form (always  in the conformal gauge),
\begin{eqnarray}\label{eqks}
u_{\tau \tau} -u_{\sigma\sigma} +{1 \o r}\left( 1 + {1 \o r}
\right)\; e^{-r}\; v\; \left[(u_{\tau})^2 -(u_{\sigma})^2\right]
- {{r\, u}\o 2}  \left[  \tea^2 -\tes^2
+\sin^2\theta\>(\pa^2-\ps^2) \right] = 0 \cr \cr
v_{\tau \tau} -v_{\sigma\sigma} +{1 \o r}\left( 1 + {1 \o r}
\right)\; e^{-r}\; u\; \left[(v_{\tau})^2 -(v_{\sigma})^2\right]
- {{r\, v}\o 2}  \left[  \tea^2 -\tes^2
+\sin^2\theta\>(\pa^2-\ps^2) \right] = 0 \; ,
\end{eqnarray}
plus eqs.(\ref{eqang}) for the angular coordinates.

The constraints  in Kruskal-Szekeres coordinates are
\begin{eqnarray}\label{vinks}
-{4\over{r}}~e^{-r}\,
\left(u_{\sigma}\;v_{\sigma}+u_{\tau}\;v_{\tau}\right) +
 r^2 \left[\tea^2+\tes^2 + \sin^2\theta\>(\pa^2+\ps^2)\right]=0\,,\cr
-{4\over{r}}~e^{-r}\,
\left(u_{\tau}\;v_{\sigma}+u_{\sigma}\;v_{\tau}\right) +
 r^2\left(\tea\tes + \sin^2\theta\>\pa\ps\right)=0 \, .
\end{eqnarray}
Notice that both the equations of motion and constraints are invariant
under the exchange $u \leftrightarrow v$.

Also notice that  the equations of motion and constraints in Kruskal-Szekeres
coordinates are regular everywhere except at the singularity $r=0$.

We shall consider closed strings where the string coordinates must be
periodic functions of $\s$:
 \begin{equation}\label{perio}
u(\s+2\pi,\tau) = u(\s,\tau) \; , \; v(\s+2\pi,\tau)= v(\s,\tau) .
\end{equation}
Therefore, the  angular coordinates $\theta,\phi$ may be just
quasiperiodic functions of $\s$:
\begin{equation}\label{qperio}
\theta(\s+2\pi,\tau) = \theta(\s,\tau) + \mbox{mod}\,2\pi\; , \;
\phi(\s+2\pi,\tau) = \phi(\s,\tau) + 2 n \pi ,
\end{equation}
where $ n$ is an integer.

\subsection {Strings Near the Singularity $r=0$}

Let us consider the solution of eqs.(\ref{eqang},\ref{eqks}) and constraints
(\ref{vinks}), near $r=0$. That is to say, near $uv=1$.

For a generic world-sheet, we choose the gauge such that   $\tau=0$
corresponds to the string at the singularity $uv=1$. This can be
achieved as shown  in general in sec. III.A.

Near the singularity $uv=1$, we propose for $ \tau\to 0 $ the expansion
\cite{negro}
\begin{eqnarray}\label{domi}
u(\s,\tau)&=& e^{a(\s)} \left[ 1 - \tau^{\a} \, \beta(\s) + \ldots
\right]\cr\cr
v(\s,\tau)&=& e^{-a(\s)} \left[ 1 - \tau^{\a'} \, {\hat \beta}(\s) + \ldots
\right]\cr\cr
\theta(\s,\tau)&=&  g(\s) +   \tau^{\lambda'}\; \mu(\s) +\ldots , \cr \cr
\phi(\s,\tau)&=& f(\s) +  \tau^{\lambda} \; \nu(\s) + \ldots \; .
\end{eqnarray}

Inserting eqs.(\ref{domi}) in  eqs.(\ref{eqang},\ref{eqks}) and constraints
(\ref{vinks}) yields\cite{negro}
\begin{eqnarray}\label{expo}
\a = \a'= 4/5 \quad &,& \quad \lambda = \lambda' = 1/5 ,\cr
 {\hat \beta}(\s) =  \beta(\s) \quad &,& \quad  \beta(\s) = {1 \o 64}
\left[\mu(\s)^2 +\nu(\s)^2 \; \sin^2 g(\s)\right]^2  \; .
\end{eqnarray}

Since the function $ \beta(\s) $ is clearly positive, we write it as
$$
 \beta(\s) = {1 \o 4}\,\gamma(\s)^4 .
$$

The coordinate $r$ then vanishes as
\begin{equation}\label{rcero}
r(\s,\tau) =\gamma(\s)^2 \;  \tau^{2/5} + \ldots \; .
\end{equation}

The string solution is completely fixed once
the functions  $ f(\s), g(\s), a(\s), \mu(\s)$ and $ \nu(\s)$ are
given. These five functions are arbitrary and can be expressed in
terms of the  initial data.

Notice that $\phi$ and $\theta$ approach their limiting values with
the same exponent $1/5$ in $\tau$.

Both the equations of motion and constraints are invariant
under the exchange $u \leftrightarrow v$ but not the boundary
conditions at $\tau = 0$. They differ by  $ a(\s) \leftrightarrow
-a(\s)$ as we see from eqs.(\ref{domi}). Therefore one can obtain
$u(\s,\tau)$ from  $v(\s,\tau)$ and
viceversa just by flipping the sign of  $ a(\s) $.

We can also find the ring solution of ref.\cite{din} setting $ f(\s)
 \equiv n \s, a(\s) \equiv 0, \mu(\s)= $ cte.  $g(\s)= $ cte.   and $
\nu(\s) \equiv 0$ [see section VIII.D].

 The corrections to  the leading behaviour
appear as positive integer powers of $\tau^{2/5}$. The subdominant
leading power in $u(\s,\tau)$ and $v(\s,\tau)$ is again
$\tau^{7/5}$. We find with the help of Mathematica\cite{negro},
\begin{eqnarray}\label{uvmas}
u(\s,\tau)&=& e^{a(\s)} \left\{ 1\right. -   \, \gamma(\s)^4 \;
\tau^{ 4/5}  \left[ 1 + O(\tau^{2/5}) \right] \cr \cr
&-& \left.  \,\gamma(\s)^6  \;
{{ f'(\sigma)\nu(\sigma)\sin^2g(\sigma)+\mu(\sigma)}\over
{28\;  a'(\sigma)}}\; \tau^{7/5} \left[ 1 + O(\tau^{2/5}) \right] \right\}
\; , \cr \cr
v(\s,\tau)&=& e^{-a(\s)} \left\{ 1\right. -  \, \gamma(\s)^4 \;
\tau^{ 4/5}  \left[ 1 + O(\tau^{2/5}) \right]  \cr \cr
&+& \left.   \,\gamma(\s)^6  \;
{{ f'(\sigma)\nu(\sigma)\sin^2g(\sigma)+\mu(\sigma)}\over
{28 \;  a'(\sigma)}}\; \tau^{7/5} \left[ 1 + O(\tau^{2/5}) \right] \right\}
\; .
\end{eqnarray}

Notice that  $u/v$ is $\tau$ independent up to order $\tau^{7/5}$.
Since $u/v = e^{-t}$, this imply that the spatial coordinate $t$ is
only $\s$-dependent up to $O(\tau^{7/5})$. More precisely,
\begin{equation}
t(\s,\tau) = \log{v\o u} = - 2 \, a(\s) +
 \,\gamma(\s)^6  \;
{{ f'(\sigma)\nu(\sigma)\sin^2g(\sigma)+\mu(\sigma)}\over
{14 \;  a'(\sigma)}}\; \tau^{7/5}   +   O(\tau^{ 9/5}) \; .
\end{equation}
In other words, $t(\s,\tau)$ varies slower than the other coordinates
$\phi$ and $r$ when the string approaches the black hole singularity
($\tau \to 0$).

Using the diagonal conformal transformation (\ref{diag}), we can fix
one of the arbitrary functions among  $ f(\s), g(\s), a(\s), \mu(\s)$
and $ \nu(\s)$ keeping in mind the periodic boundary conditions:
\begin{eqnarray}
 a(\s+ 2\pi ) =  a(\s)\; , \; \nu(\s + 2\pi )= \nu(\s)\; , \; \mu(\s +
2\pi )= \mu(\s)\; , \cr \cr
\; f(\s+ 2\pi)=f(\s) + 2 n \pi \; , \;  g(\s+2\pi)= g(\s) \; \mbox{mod}\;  2
\pi \; .
\end{eqnarray}

We are left with {\bf four} arbitrary functions of $\s$. This is
precisely the number of transverse string degrees of freedom.

\subsection{String energy-momentum  and  invariant size near the
singularity}

The string size in the Schwarzschild metric takes the form
\begin{equation}\label{tama}
S^2  =   G_{AB}(X) \, {\dot X}^A \, {\dot X}^B = {4 \o r}\, e^{-r}
{\dot u}{\dot v} -   r^2
\,  {\dot\theta}^2 - r^2 \, {\dot\phi}^2 \, \sin^2\theta .
\end{equation}
where we used eqs.(\ref{TC}) and (\ref{schsch}).

We find near the singularity at $r=0$ using eqs.(\ref{rcero}-\ref{uvmas})
\begin{eqnarray}\label{tamaO}
S &=& {{ 4\,  {a'(\s)}^2}\o { \gamma(\s)^2}} \; \tau^{ -2/5} -
{4 \o 7}\; a'(\s)^2 \; \left( 6 + 25 \; {{ a'(\s)^2}\o
{\gamma(\s)^7}} \right) + O(\tau^{2/5}) \cr \cr
&=& {{ 4\,  {a'(\s)}^2}\o r}  -
{4 \o 7}\; a'(\s)^2 \; \left( 6 + 25 \; {{ a'(\s)^2}\o
{\gamma(\s)^7}} \right) + O(r) \; .
\end{eqnarray}
For simplicity we choose here an equatorial solution at $\theta = \pi/2$.
The invariant string size tends then to infinite when the string falls
into the   $r=0$ singularity\cite{negro,nulbh}. This is due to the
infinitely growing
gravitational forces that act there on the string.

The string stretching near  $r=0$ was first observed in ref.\cite{cl} using
perturbative methods and in ref.\cite{bhln} for a family of exact string
solutions inside the horizon.

\bigskip

Inside the horizon we can use  $t, \theta, \phi$
as spatial coordinates and $r$ as a coordinate time. We find,
\begin{equation}\label{Teta}
\sqrt{g_{rr}}\,\Theta^{AB}(r)  =  \frac{1}{2\pi \alpha'} \int d\s d\tau
\left( {\dot X}^A {\dot X}^B -X'^A X'^B \right)
\delta(r - r(\tau,\s ) ).
\end{equation}
where $g_{rr} = r/(1-r) > 0$.

We have for the black hole case:
\begin{equation}
G_{AB}(X) \;  {\dot X}^A {\dot X}^B = {{r {\dot r}^2}\o{1-r}} - r^2
\,  {\dot\theta}^2 - r^2 \, {\dot\phi}^2 \, \sin^2\theta - {{1-r}\o
r}\, {\dot t}^2 \; .
\end{equation}
Using eqs.(\ref{domi}) and (\ref{rcero}) for $\tau \to 0$, we find
that each of the first three terms grows as $\tau^{-4/5}$
whereas the last term vanishes as  $\tau^{2/5}$. Moreover,  the sum
of the three terms $O(\tau^{-4/5})$ identically vanishes thanks to
eq.(\ref{expo}). This cancellation in the trace tells us that near $r=0$, the
dominant (and divergent) components $T_r^r, T_{\phi}^{\phi}$ and
$T_{\theta}^{\theta}$ yield a zero trace. This means that the string
behaves to leading order as  {\bf two}-dimensional massless
particles\cite{negro}. This is the so-called dual to unstable
behaviour \cite{erice}  (here for two spatial dimension).

For $\tau \to 0, \; r \to 0$ we can use in eq.(\ref{Teta}) the dominant
behaviours:
\begin{eqnarray}
r(\s,\tau) &=&\gamma(\s)^2 \;  \tau^{2/5} + O( \tau^{4/5}) , \cr \cr
\theta(\s,\tau)&=&  g(\s) +  \mu(\s) \, \tau^{1/5} + O(
\tau^{3/5})  , \cr \cr
\phi(\s,\tau)&=&   f(\s) +  \, \nu(\s)  \, \tau^{1/5} + O(
\tau^{3/5})  , \cr \cr
t(\s,\tau)& =& - 2 a(\s)  +
 \gamma(\s)^6 \;
{{f'(\sigma)\nu(\sigma)\sin^2g(\sigma)+\mu(\sigma)}\over
{14 \;  a'(\sigma)}}\; \tau^{7/5}
+ O\left(\tau^{9/5}\right)\; .
\end{eqnarray}
We thus find for $ r \to 0 $,
\begin{eqnarray}
2\pi\alpha'\,\Theta^{rr}(r) & = & {2\o {5\;r^2}}
\int_0^{2\pi}{\rm d}\s\, |\gamma(\s)|^5 + O({1 \o r})
\to +\infty \,,\cr\cr
2\pi\alpha'\,\Theta^{\phi\phi}(r) & = &  {1\o {10\;r^3}}
\int_0^{2\pi}{\rm d}\s\, \nu(\s)^2 \; |\gamma(\s)|^3
+ O({1 \o {r^2}}) \to +\infty\,,\cr\cr
2\pi\alpha'\,\Theta^{\theta\theta}(r) & = &  {1\o {10\;r^3}}
\int_0^{2\pi}{\rm d}\s\, \mu(\s)^2 \; |\gamma(\s)|^3 + O({1 \o
{r^2}}) \to +\infty \,,\cr\cr
2\pi\alpha'\,\Theta^{tt}(r) & = & -10\;r
\int_0^{2\pi}{\rm d}\s\,{{\left[a'(\s)\right]^2}\o{ |\gamma(\s)|^5
}}+ O(r^2)\to 0^-
\, \, .
\end{eqnarray}

We can identify the string energy with the mixed component
$-\Theta_r^r$. We define the mixed components $\Theta_A^B(r)$
by integrating $T_A^B(X)$ over the spatial volume.

This yields for $r\to 0$,
\begin{equation}\label{ener}
E \equiv -\Theta_r^r = { 1 \o {2\pi\alpha'}}\;
 {2\o {5\;r}}
\int_0^{2\pi}{\rm d}\s\, |\gamma(\s)|^5 + O(1) \to +\infty \; .
\end{equation}

The transverse pressures are defined as the mixed components
$\Theta_{\phi}^{\phi}$ and $\Theta_{\theta}^{\theta}$. They diverge
for $r\to 0 $ :
\begin{eqnarray}\label{pres}
P_{\phi} \equiv \Theta_{\phi}^{\phi} &=&  { 1 \o {2\pi\alpha'}}\;  {2\o {5\;r}}
\int_0^{2\pi}{\rm d}\s\, \nu(\s)^2 \;  \sin^2 g(\s) \; |\gamma(\s)|^5
\,\to +\infty\,,\cr\cr
P_{\theta} \equiv \Theta_{\theta}^{\theta} &=&  { 1 \o {2\pi\alpha'}}\;
{2\o {5\;r}}
\int_0^{2\pi}{\rm d}\s\, \mu(\s)^2 \; |\gamma(\s)|^5 \,\to +\infty .
\end{eqnarray}
Thus, to leading order,
$$
E = P_{\theta}+ P_{\phi} \quad \mbox{for}\quad r \to 0 \; .
$$
exhibiting a two-dimensional ultrarelativistic gas behaviour.
The tidal forces  infinitely stretch the
string near $r=0$ in effectively only two directions: $\phi$ and $\theta$.

We find for the off-diagonal components,
\begin{eqnarray}
2\pi\alpha'\,\Theta^{t r}(r) &=& {{r^{1/2}}\over 10}\; \int_0^{2\pi}{{{\rm
d}\s\,}\o{a'(\s)}}\, \gamma(\s)^4\;
\left[f'(\sigma)\nu(\sigma)\sin^2g(\sigma)+\mu(\sigma)\right]\to 0^+
\; , \cr\cr
2\pi\alpha'\,\Theta^{t\theta}(r) &=& \, {2\over 5}\; \int_0^{2\pi}{\rm
d}\s\,{{\mu(\s)}\o{a'(\s)}}\, |\gamma(\s)|^3 \;
\left[f'(\sigma)\nu(\sigma)\sin^2g(\sigma)+\mu(\sigma)\right] = O(1)\; , \cr\cr
2\pi\alpha'\,\Theta^{t\phi}(r)  &=& \, {2\over 5}\; \int_0^{2\pi}{\rm
d}\s\,{{\nu(\s)}\o{a'(\s)}}\, |\gamma(\s)|^3 \;
\left[f'(\sigma)\nu(\sigma)\sin^2g(\sigma)+\mu(\sigma)\right] = O(1)\; , \cr\cr
2\pi\alpha'\,\Theta^{r \phi }(r)  &=&\, {1\o {5\;r^{5/2}}}
\int_0^{2\pi}{\rm d}\s\, \nu(\s)\,\gamma(\s)^4 \,\to \infty , \cr\cr
2\pi\alpha'\,\Theta^{r \theta}(r)  &=&\, {1\o {5\;r^{5/2}}}
\int_0^{2\pi}{\rm d}\s\, \mu(\s)\, \gamma(\s)^4\;
\,\to \infty  , \cr\cr
2\pi\alpha'\,\Theta^{\theta\phi}(r)  &=&\, {1\o {10 \;r^3}}
\int_0^{2\pi}{\rm d}\s\,  \mu(\s)\, \nu(\s)\, |\gamma(\s)|^3\; \to \infty \; .
\end{eqnarray}

Notice that the invariant string size tends to infinity [see
eq.(\ref{tamaO})] with $4\, a'(\s)^2$ as proportionality factor. Since
$-2 a(\s)$ is the leading behaviour of $t(\s,\tau)$, this suggests us that
the string stretches  infinitely in the (spatial) $t$ direction when $r\to 0$.

As a matter of fact, infinitely growing string sizes are not observed in
cosmological spacetimes  \cite{erice,cos} for strings exhibiting radiation
 (dual to unstable) behaviour.

For particular string solutions the energy-momentum tensor and the
string size can be less singular than in the generic case.
For ring solutions \cite{din},  $\mu(\s) = \mu =$ constant, $g(\s) =
g =$  constant, $a(\s) = \nu(\s) = 0$,
there is no stretching and
$$
S = r \sin{g} \to 0
$$
$$
E =  P_{\theta} =  { 1 \o {\alpha'}}\;{{\mu^5}\o {80\;r}} \,\to
+\infty
\quad , \quad P_{\phi} = 0 \; .
$$
There is no string stretching but the string keeps exhibiting dual to
unstable behaviour. This is due to the balance of the tidal forces
thanks to the special symmetry  of the solution.  It behaves in this
special case as {\bf one}-dimensional massless particles for $r \to 0$.

As is easy to see, setting $\mu(\s) = 0, g(\s) = \pi/2$
all equatorial string solutions   behave as {\bf
one}-dimensional massless particles for $r \to 0$.

The resolution method used here for strings in
black hole spacetimes is analogous  to the expansions for $\tau\to 0$
developped  in ref.\cite{sv,gsv} for  strings in cosmological
spacetimes (see sec. IV.A).

\subsection{Axisymmetric Ring Solutions}

We present in this section an axisymmetric ansatz describing a ring
string in Schwarzschild spacetime. This  ansatz has a symmetry
compatible with the  equations (\ref{eqsch}, \ref{eqang}) and
(\ref{vinsch}) and hence separates them
into ordinary differential equations \cite{din}.
\begin{equation}\label{anbh}
\phi=n\sigma\,,\quad\theta=\theta(\tau)\,,\quad t=t(\tau)\,,\quad
r=r(\tau)\,.
\end{equation}
where $n =$ integer. This ansatz inserted in equations  (\ref{eqsch},
\ref{eqang}) and (\ref{vinsch}) produces the following set of equations:
\begin{eqnarray}\label{anbeq}
\ddot r - (r-3/2)\; \dot\theta^2 + n^2\sin^2\theta\; (r-1/2)&=0\,,\cr
r\ddot\theta + 2\dot r\dot\theta +n^2 r\sin\theta \cos\theta&=0\,.
\end{eqnarray}
with a conserved quantity
\begin{equation}\label{anbco}
e^2=\dot r^2+r(r-1)(\dot\theta^2+n^2\sin^2\theta)\,,
\end{equation}
and $t$ given by
\begin{equation}\label{anbtd}
\dot t={{er}\over{r-1}}\,.
\end{equation}
Note that these equations are invariant under the
change $\tau\to-\tau$, $t\to-t$.
Since $\dot t > 0$ outside the horizon, $t(\tau)$ is a
monotonous function, and we can use
either $\tau $ or $t$ to study the time evolution for $r > 1$.

The string energy  is found to be in this case,
$$
E(t)\equiv -P_0 = -{{G_{00}}\o{\a'}}{{dX^0}\o{d\tau}} =
{e\over{\a'}}\,.
$$
where we used eq.(\ref{tens}):
$$
P^0 = \int d^{D-1}X \sqrt{-G}~ T^{00}(X) \; .
$$

The invariant size of the string in this case is
$$
S=n r \sin\theta \, .
$$
A useful equation, satisfied by solutions of these equations, is
\begin{equation}\label{osceq}
\left({{{\rm d}^2}\over{{\rm d}\tau^2}}+ n^2\right)(r\sin\theta)=
{1\over2}(n^2\sin^2\theta - 3\dot\theta^2)\sin\theta\,.
\end{equation}

Let us now examine the possible asymptotic behaviours of these
equations in different regimes. A first interesting question
is the existence of collapsing solutions and the corresponding
critical exponents.
On computation, we find two possible collapsing behaviours {}from
(\ref{anbeq}-\ref{anbco})
, with the adequate choice of origin for $\tau$
for  $\tau \to 0$:
\begin{eqnarray}\label{caidau}
r&\buildrel{{\tau\to 0}}\over\simeq&\alpha \; \tau^{2/5}\,,\cr
\theta&\buildrel{{\tau\to 0}}\over\simeq&
\theta_f+ 2\; \sqrt{\alpha} \; \tau^{1/5}\,,
\end{eqnarray}
where $\a $ is a constant and
\begin{eqnarray}
r&\buildrel{{\tau\to 0}}\over\simeq &~e\tau +
 {{n^2 \sin^2\theta_f}\over4}~\tau^2 -
{{n^2 e\sin^2\theta_f}\over6}~\tau^3 + O(\tau^4)\,,\cr
\theta&\buildrel{{\tau\to 0}}\over\simeq &~\theta_f -
{{n^2 \sin(2\theta_f)}\over{12}}~\tau^2
+{{n^2\sin^3\theta_f\cos\theta_f}\over{726}}~\tau^3+ O(\tau^4)\,.
\end{eqnarray}
This last one is obviously subdominant with respect to eq.(\ref{caidau}) .

Consider now the regime given by large $ r $ and $ \dot\theta^2/r $ small.
{}From equations (\ref{anbco}) and (\ref{osceq}), we have that, for
$|\tau|\to+\infty$,
\begin{eqnarray}\label{bhinfas}
r&\buildrel{{\tau\to +\infty}}\over\simeq & p|\tau|\,,\cr
\theta&\buildrel{{\tau\to +\infty}}\over\simeq & \theta_0 -
{{m}\over{p}}{{\cos(n\tau+\varphi_0)}\over\tau}\,,
\end{eqnarray}
together with
$$
e^2=p^2+n^2 m^2\,,
$$
and $t\buildrel{{\tau\to +\infty}}\over\sim e\tau$.
Here $\theta_0$ is such that $\sin\theta_0=0$,
i.e., $\theta=l\pi$ with $l$ an integer.
We could here understand $p$ as an asymptotic radial momentum, $e$
the energy, and
$n m$ the mass of the string. The latter is determined by the amplitude
of the string oscillations.

We find for large $\tau$ that
\begin{eqnarray}\label{coorxy}
x=r\sin\theta \cos\phi& = (-1)^{l+1} ~m~ \cos(n\tau+\varphi_0)
\cos{n\sigma} \,,\cr
y=r\sin\theta \sin\phi&
= (-1)^{l+1} ~m~\cos(n\tau+\varphi_0)\sin n\sigma\,.
\end{eqnarray}
In this region, spacetime is minkowskian, and we can recognize
(\ref{coorxy}) as
the $n^{{\rm th}}$ excitation mode of a closed string. For $|n|=1$ this
corresponds at the quantum level to a graviton and/or a dilaton.
Notice that $m$ is  the amplitude of the string oscillations.

The string size is here $ S(\tau) =  r(\tau) ~ |\sin\theta(\tau)|
$. We find from eqs.(\ref{caidau} -\ref{bhinfas}) that
\begin{eqnarray}
S(\tau)&\buildrel{{\tau\to +\infty}}\over\simeq &  m |\cos(\tau +
\varphi)|\buildrel{{\tau\to +\infty}}\over\sim {{ m}\o {\sqrt2}}\cr
S(\tau)&\buildrel{{\tau\to +0}}\over\simeq &  \a
\sin\theta_f~\tau^{2/5} \; .
\end{eqnarray}

 Whenever the string is not swallowed by the black hole, equation
(\ref{bhinfas}) describes both the incoming and outgoing regions
$\tau\to\pm\infty$. However, the mass $m$, the momentum $p$ and
the phase $\varphi_0$ are in general different in the two
asymptotic regions. This is an illustration of a rather
general phenomenon noticed at the quantum level: particle
transmutation \cite{tras}. This means that the excitation
state of a string changes in general when it is scattered
by an external field  like a black hole. Within our
classical ansatz (\ref{anbh}), the only possible changes are in
amplitude (mass), momentum, and phase.
It can be seen numerically that the excitation state is
indeed modified by the interaction with the black hole.

Due to the structure of our ansatz, the string, if it is not
absorbed for some finite $\tau$, may return to $z=+\infty$
($\theta_f=0\pmod{2\pi}$), where it started at $\tau=-\infty$,
or go past the black hole towards $z=-\infty$ ($\theta_f=\pi\pmod{2\pi}$)

An special case of interest is that of solutions such that
$r(\tau)=r(-\tau)$. It follows that
$\dot\theta^2(\tau)=\dot\theta^2(-\tau)$, from which
$\theta(\tau)=\Delta-\epsilon\theta(-\tau)$,
with $\epsilon$ a sign. We
then see that $\Delta$ is restricted to multiples of
$\pi$ if the string does not fall into the black hole.
If it is an odd multiple, we understand that
the string has circled round the black hole a number
of times and then  has continued to infinity, whereas
when it is an even multiple, the string bounces back
after some dithering around the black hole. This analysis can
be extended to all solutions.

Let us now analyze the absorption of the string by the
black hole. If $r$ starts at $+\infty$ for $\tau=-\infty$ and
decreases ($\dot r<0$), $\dot r$ must change sign at the
periastron at time $\tau_0$. Otherwise the singularity at
$r=0$ will be reached. Furthermore, $\ddot r(\tau_0)>0$.
We see {}from the first equation in the set (\ref{anbeq}) that
this implies that $r(\tau_0)>3/2$. In other words, if the
string  penetrates the $r<3/2$ region, it will necessarily
fall into the singularity. In yet another paraphrase,
there is an effective horizon
for ring string solutions. The surface $r=3/2$ is necessarily
contained within this horizon. Let us recall that for massless
geodesics the  effective horizon is a sphere of radius
$r = {3 \over 2} \sqrt3 $.

To illustrate these points, we adjunct some figures.
They depict the motion of ring-like strings described by
equations (\ref{anbh}) through (\ref{anbtd}). We numerically integrate
equations (\ref{anbeq}) {}from large negative $\tau$, where the
asymptotic behaviour (\ref{bhinfas}) holds. We choose $\theta_0$,
$n=1$, and vary the values of $p,m$ and $\varphi_0$. Depending on
this last set of three values  the string is absorbed or not
by the black hole.

In figs. 8 we show an example of direct fall (i.e.,
with no bobbing around the black hole).

\begin{center}
Figs. 8~:
Numerical solution for the equations of motion of a string
in a Schwarzschild black hole background: the string falls into the
black hole.
\end{center}

In order to
compare with this one, we next portray (figs. 9) a case where the string
goes past the black hole before returning to it and collapsing. The
clearest view of this event is given by fig. figbhpas d, which depicts
$z=r\cos\theta$ as a function of  $\rho=r\sin\theta$.

\begin{center}
Figs. 9~: Numerical solution for the equations of motion of a string
in a Schwarzschild black hole background: the string falls into the
black hole, but only after first going past it and then back into
the singularity.
\end{center}

Figs. 10 is dedicated to a non-falling string. It is particularly
interesting to point out that the excitation state has been
changed by scattering by the black hole, as can be clearly seen {}from
the third graph in this figure, which depicts $r\sin\theta$ as a function
of $\tau$.
We see that the oscillation amplitude is larger after the collision
than before. This means that the outgoing string mass is larger than the
ingoing string mass. Hence, particle transmutation in the sense
of ref.\cite{tras} takes place here.

In the fourth of this series, fig. 10 d,
we portray $z=r\cos\theta$ against $\rho=r\sin\theta$. It is to be remarked
that the string {\bf bounces} (the lower end of the picture), then oscillates
around the black hole, and finally escapes to infinity.

\begin{center}
Figs. 10~: Numerical solution for the equations of motion of a string
in a Schwarzschild black hole background: the string goes past the black
hole, circles round it, and then bounces back with
a change in its amplitude and momentum.
\end{center}

That the string be absorbed or not by the black hole is dictated by
whether it comes or not within the effective horizon,
as mentioned above. This, in turn, is crucially dependent on the phase
$\varphi_0$ chosen as part of the initial data ($\tau\to-\infty$).
Whatever value the mass (amplitude) $m$ and the momentum $p$ take,
there is always some interval of values of $\varphi_0$ for which
the string will be absorbed by the black hole.

Besides numerical experiments, this behaviour follows {}from the simple
fact that a change of the initial phase $\varphi_0$ would displace the
string worldsheet, thus possibly bringing it closer to the black hole.

\newpage

{\bf Figure Captions}:

\bigskip

{\bf Figure 1:} Plot of the function $H T(\tau)$, for two
values of $\sigma$, for $n = 4, |x^0> = (1+i,.6+.4i,.3+.5i,.77+.79i)$.
The function $\tau(T)$ is multivalued,
revealing the presence of five strings.

\medskip

{\bf Figure 2:} Same as fig.1, for $n = 4, |x^0> = (1,-1,i,1). $
Because of a degeneracy, there are now only three strings.

\medskip

{\bf Figure 3:} $\tau = \tau(\sigma,T)$ for fixed $T$
for $n = 4, |x^0>=(1,-1,i,1)$. Three values of HT are
displayed, corresponding to HT=0 (full line), 1
(dots), and 2 (dashed line). For each HT, three curves are
plotted, which correspond to the three strings. They are ordered with
$\tau$ increasing.

\medskip

{\bf Figure 4:} Same as fig. 3 for
$n = 4, x^0>=(1+i,.6+.4i,.3+.5i,.77+.79i).$
a) The five curves corresponding to the five strings at HT=2.
b) The five curves for three values of HT: HT=0 (full line), 1
(dots), and 2 (dashed line).
\medskip

{\bf Figure 5:} Evolution as a function of cosmic time $HT$ of the
three strings, in the comoving coordinates $(X^1,X^2)$,
for $n = 4, |x^0> = (1,-1,i,1)$.
The comoving size of string (1) stays constant
for $H T <-3$, then decreases
around $H T = 0$, and stays constant again after $HT = 1$.
The invariant size of string (2) is constant for negative $HT$,
then grows as the expansion factor  for $HT > 1$,
and becomes identical to string (1).
The string (3) has a constant comoving size for $ HT < -3$,
then  collapses as $e^{-HT}$ for positive $HT$.

\medskip

{\bf Figure 6:} Evolution of three of the five strings for
$ n = 4, |x^0> = (1+i,.6+.4i,.3+.5i,.77+.79i)$.

\medskip

{\bf Figure 7:} Evolution of the three strings
for the degenerate case $n=6, |x^0> = (1,-1,i,1)$.

\bigskip

{\bf Note:

Figures 1 to 7 can be find in ref.\cite{cdms}.

Figures 8 to 10  can be find in ref.\cite{din}}


\begin{thebibliography}{11}

\bigskip


\bibitem{erice} Lectures by
H. J. de Vega and N. S\'{a}nchez in `String Quantum
 Gravity and the Physics at the
              Planck Scale', Proceedings of the Erice Workshop held in June
              1992. Edited by N. S\'{a}nchez, World Scientific, 1993.
              Pages 73-185, and references given therein.

 Lectures by H. J. de Vega and N. S\'{a}nchez in `Current Topics in
Astrofundamental Physics: The Early Universe',  Proceedings of the
Nato ASI Third D. Chalonge School, 4-16 September 1994, p. 99-128,  edited by
N. ~S\'anchez and A. Zichichi, Kluwer, 1995.

\bibitem{dvs87} H. J. de Vega and N. S\'anchez, Phys. Lett. {\bf B 197}, 320
(1987).

\bibitem{cos} H. J. de Vega and N. S\'anchez,
Phys. Rev. {\bf D50}, 7202 (1994).

\bibitem{prd} H. J. de Vega and N. S\'anchez,
Phys. Rev. {\bf D47}, 3394 (1993).

\bibitem{dms} H. J. de Vega, A. V. Mikhailov and N. S\'{a}nchez, Teor. Mat.
              Fiz. {\bf 94} (1993) 232.

\bibitem{cdms} F. Combes, H. J. de Vega, A. V. Mikhailov and
N. S\'{a}nchez,

Phys. Rev. {\bf D50}, 2754 (1994).

\bibitem{dls} H. J. de Vega, A. L. Larsen and N. S\'anchez,
Nucl. Phys. {\bf B 427}, 643 (1994).

\bibitem{igor} I. Krichever, Funct. Anal. and Appl. {\bf 28}, 21 (1994),

[Funkts. Anal. Prilozhen.  {\bf 28}, 26 (1994)].

\bibitem{agn} H. J. de Vega and N. S\'anchez,
Nucl. Phys. {\bf B309}, 552 and 577 (1988).

\bibitem{sv} N. S\'anchez and G. Veneziano,
Nucl. Phys. {\bf B333}, 253 (1990).

\bibitem{gsv} M. Gasperini, N. S\'anchez and G. Veneziano,

Int. J. Mod. Phys. {\bf A 6},  3853 (1991) and
Nucl. Phys. {\bf B364}, 365 (1991).

\bibitem{din} H. J. de Vega and I. L. Egusquiza,
Phys. Rev. {\bf D49}, 763 (1994).

\bibitem{ads}  A. L. Larsen and N. S\'anchez,
Phys. Rev. {\bf D50}, 7493 (1994).

\bibitem{bhln} A. L. Larsen and N. S\'anchez,
Phys. Rev. {\bf D51}, 6929 (1995).

\bibitem{multi} H. J. de Vega and I. L. Egusquiza,

 hep-th/9505029, submitted to Class. and Quantum Grav.

\bibitem{hawel} S. Hawking and G. F. R. Ellis, `The large scale
structure of the spacetime',

Cambridge Univ. Press, 1973.

\bibitem{ijm} H. J. de Vega and N. S\'anchez,
Int. J. Mod. Phys. {\bf A 7}, 3043 (1992).

\bibitem{negro}   H. J. de Vega and I. L. Egusquiza,
`Strings next and inside black holes',

hep-th/9506214, to appear in Phys. Rev. D.

\bibitem{ell} G. F. R. Ellis, Banff Lectures 1990, {\it in} Gravitation,

eds. R. Mann and P. Wesson , World Scientific 1991.

\bibitem{twbk}See for a review, T. W. B. Kibble, Erice Lectures at the
Chalonge School

in Astrofundamental Physics, N. S\'anchez editor, World Scientific, 1992.

\bibitem{vil} A. Vilenkin, Phys. Rev. {\bf D 24},  2082 (1981) and
Phys. Rep. {\bf 121}, 263 (1985).

N. Turok and P. Bhattacharjee, Phys. Rev. {\bf D 29}, 1557 (1984).

\bibitem{nic}  H. J. de Vega and A. Nicolaidis,
	Phys. Lett. {\bf B 295}, 214 (1992).

  H. J. de Vega, I. Giannakis and  A. Nicolaidis,
	Mod. Phys. Lett. {\bf A 10}, 2432 (1995).

\bibitem{nulbh}  C. Loust\'o and N. S\'anchez,
`String dynamics in cosmological and

black hole backgrounds: the null string approach', in preparation.

\bibitem{tse} R.~Myers,  Phys. Lett.   {\bf B199},  371 (1987).

M.~Mueller,  Nucl. Phys.   {\bf B337},  37 (1990).

See for a review: A.A.~Tseytlin in the Proceedings
of the Erice School

``String Quantum Gravity and Physics at the Planck Energy Scale'',

21-28 June 1992, Edited by N. ~S\'anchez, World Scientific, 1993.

\bibitem{wei} S. Weinberg, `Gravitation and Cosmology', J. Wiley, 1972.

\bibitem{romi} E. W. Kolb and M. S. Turner, `The Early Universe',
Addison-Wesley, 1990.

\bibitem{linde} A. D. Linde, `Particle Physics and Inflationary
Cosmology', Harwood (1990).

\bibitem{eqef}  See for example,

I. Antoniadis, C. Bachas, J. Ellis and D. V. Nanopoulos,

 Nucl. Phys. {\bf B 328}, 117 (1989) and  Phys. Lett.  {\bf B 257}, 278 (1991).


B. A. Campbell, A. Linde and K. A. Olive,
 Nucl. Phys. {\bf B 355}, 146 (1991).

 B. A. Campbell, N. Kaloper and K. A. Olive,
 Phys. Lett.   {\bf B 277}, 265 (1992).

A.A.~Tseytlin, Mod. Phys. Lett {\bf A 6}, 1721 (1991).

A.A.~Tseytlin and C. Vafa, Nucl. Phys. {\bf B 372}, 443 (1992).

R. Brustein and P. J. Steinhardt, Phys. Lett. {\bf B 302}, 196 (1993).

M. Gasperini and G. Veneziano, Mod. Phys. Lett. {\bf A 8}, 370 (1993),

Phys. Lett.   {\bf B 277}, 256 (1992)
and Astroparticle Physics {\bf 1}, 317 (1993).

E. Raiten, Nucl. Phys. {\bf B 416}, 881 (1994).

R. Brustein and  G. Veneziano, Phys. Lett. {\bf B 329}, 429 (1994).

V. A. Kosteleck\'y and M. J. Perry,  Nucl. Phys. {\bf B 414}, 174 (1994).

See in addition ref.\cite{tse}.

\bibitem{gsw} See for example: \\
M. Green, J. Schwarz, E. Witten,  `Superstring Theory'.

Cambridge University Press. 1987.

\bibitem{grad}I.S. Gradshteyn and I.M. Ryzhik,
Table of Integrals Series and Products,

(Academic Press, New York, fourth edition, 1965).

\bibitem{ll}See for example,

S. Weinberg, `Gravitation and Cosmology', J. Wiley, 1972.

\bibitem{tur} F. M\"uller-Holstein, Class. Quant. Grav. {\bf 3}, 665 (1986).

K. G. Akdeniz et al. Mod. Phys. Lett. {\bf A 6}, 1543 (1991) and

Phys. Lett. {\bf B 321}, 329 (1994).

\bibitem{fhh} M. V. Fischetti, J. B. Hartle and B. L. Hu, Phys. Rev.
{\bf D 20}, 1757 (1979).

\bibitem{uno} S. Wada and T. Azuma, Phys. Lett. {\bf B 132}, 313 (1983).

V. Sahni and L. A. Kofman, Phys. Lett. {\bf A 117}, 275 (1986).

\bibitem{dos} M. A. Castagnino, J. P. Paz and  N. S\'anchez,
Phys. Lett. {\bf B 193}, 13 (1987).

\bibitem{tres} D. G. Boulware and S. Deser,
Phys. Rev. Lett. {\bf  55}, 2656 (1985).

\bibitem{zm1}V. E. Zakharov and A. V. Mikhailov, JETP, {\bf 75}, 1953 (1978).

\bibitem{dv78} H. J. de Vega,  Phys. Lett. {\bf B 87}, 233 (1979).

\bibitem{ist} C. S. Gardner, J. M. Greene, M. D. Kruskal and R. M. Miura,

Phys. Rev. Lett. {\bf 19}, 1095 (1967).

P. D. Lax, Comm. Pure and Appl. Math. {\bf 21} 467 (1968).

\bibitem{sol} M. J. Ablowitz and H. Segur,
``Solitons and the Inverse scattering
transformation'',

SIAM Philadelphia 1981.

V.E. Zakharov, S.V. Manakov, S.P. Novikov and L.P. Pitaevsky,

``Soliton Theory; The Inverse Method'', Nauka, Moscow, 1980.

\bibitem{lib2} A. C.  Scott, F.  Y. F. Chu and D. W. MacLaughlin,
Proc. IEEE, {\bf 61}, 1443 (1973).

G. L. Lamb, Elements of Soliton Theory, J. Wiley, NY (1980).

\bibitem{cl} C. Loust\'o and N. S\'anchez, Phys. Rev. {\bf D47}, 4498 (1993).
\bibitem{bibneg} See for example,

C. W. Misner, K. S. Thorne and J. A. Wheeler, `Gravitation', Freeman, 1973.

 \bibitem{ondch} H. J. de Vega and N. S\'anchez, Nucl. Phys. {\bf B 317},
706 (1989) .

 D. Amati and K. Klim\^cik, Phys. Lett. {\bf B 210} , 92 (1988)  ,

 see also: ref.\cite{costa}.

 \bibitem{costa} M. Costa
 and H. J. de Vega, Ann. Phys. {\bf 211}, 223 and 235 (1991).
  \bibitem{ondpl} H. J. de Vega and N. S\'anchez, Phys. Rev. {D 45} ,
2783 (1992).

 H. J. de Vega, M. Ram\'on Medrano and N. S\'anchez,

  Class. and Quantum Grav. {\bf 10}, 2007 (1993).

G. Horowitz and A.R. Steif, Phys. Rev. Lett. {\bf 64}, 260 (1990) and

 Phys. Rev. {\bf D 42} , 1950 (1990).
 \bibitem{conico} H. J. de Vega and N. S\'anchez,  Phys. Rev. {\bf D 42}, 3969
(1990) and

 H. J. de Vega, M. Ram\'on Medrano and N. S\'anchez, Nucl. Phys. {\bf B
374}, 405 (1992).

\bibitem{tras} H.J.~de~Vega, M.~Ram\'on~Medrano and N.~S\'anchez,
 Nucl. Phys.  {\bf B351}, 277 (1991).


\end{thebibliography}
\end{document}